\documentclass[letterpaper]{article} 
\usepackage[draft]{aaai25}  
\usepackage{times}  
\usepackage{helvet}  
\usepackage{courier}  
\usepackage[hyphens]{url}  
\usepackage{graphicx} 
\usepackage{natbib}  
\usepackage{caption} 
\usepackage{booktabs}       
\usepackage{microtype}      
\usepackage{xcolor}         

\usepackage{soul}
\usepackage{url}
\usepackage{amsmath}
\usepackage{amsthm}

\usepackage{xspace,tabularx,multirow}
\usepackage{tikz}
\usepackage{pgfplots}
\pgfplotsset{compat=1.16}
\usetikzlibrary{patterns}
\usepackage{subfig}
\usepackage{colortbl}
\usepackage{bbold}
\usepackage{enumitem}
\usepackage{tablefootnote}
\usepackage{upgreek,textgreek}
\usepackage{titlecaps}
\usepackage{lipsum}
\usepackage{bbding}

\usepackage[ruled, vlined, linesnumbered]{algorithm2e}

\usepackage{newfloat}
\usepackage{listings}

\DeclareCaptionStyle{ruled}{labelfont=normalfont,labelsep=colon,strut=off} 
\newcommand{\renchi}[1]{{\color{red}{[Renchi: #1]}}}

\makeatletter
\newcommand*\bigcdot{\mathpalette\bigcdot@{.5}}
\newcommand*\bigcdot@[2]{\mathbin{\vcenter{\hbox{\scalebox{#2}{$\m@th#1\bullet$}}}}}
\makeatother

\newcommand{\stitle}[1]{\vspace*{0.5em}\noindent{\bf #1.\/}}

\newcommand{\V}{{V}\xspace}
\newcommand{\G}{{G}\xspace}
\newcommand{\N}{{N}\xspace}

\newcommand{\EDG}{{E}\xspace}

\newcommand{\AM}{\mathbf{A}\xspace}

\newcommand{\DM}{\mathbf{D}\xspace}
\newcommand{\IM}{\mathbf{I}\xspace}

\newcommand{\PM}{\mathbf{P}\xspace}

\newcommand{\LM}{\mathbf{L}\xspace}
\newcommand{\UM}{\mathbf{U}\xspace}
\newcommand{\VM}{\mathbf{V}\xspace}
\newcommand{\HM}{\mathbf{H}\xspace}

\newcommand{\BM}{\mathbf{B}\xspace}

\newcommand{\EM}{\mathbf{E}\xspace}

\newcommand{\xm}{\mathbf{x}\xspace}
\newcommand{\zm}{\mathbf{z}\xspace}
\newcommand{\thetam}{\boldsymbol{\gamma}\xspace}
\newcommand{\ym}{\mathbf{y}\xspace}
\newcommand{\sm}{\mathbf{s}\xspace}

\newcommand{\eat}[1]{}

\newcommand{\algo}{ECHO\xspace}
\newcommand{\algoabbr}{ECHO\xspace}

\newenvironment{customlegend}[1][]{%
    \begingroup
    \csname pgfplots@init@cleared@structures\endcsname
    \pgfplotsset{#1}%
}{%
    \csname pgfplots@createlegend\endcsname
    \endgroup
}%

\def\addlegendimage{\csname pgfplots@addlegendimage\endcsname}

\makeatletter
\newcommand\footnoteref[1]{\protected@xdef\@thefnmark{\ref{#1}}\@footnotemark}
\makeatother

\let\oldnl\nl
\newcommand{\nonl}{\renewcommand{\nl}{\let\nl\oldnl}}

\definecolor{myred}{HTML}{fd7f6f}
\definecolor{myred_new}{HTML}{D8D8D8}
\definecolor{myred_new2}{HTML}{D7191C}
\definecolor{myblue}{HTML}{7eb0d5}
\definecolor{mygreen}{HTML}{b2e061}
\definecolor{mypurple}{HTML}{bd7ebe}
\definecolor{myorange}{HTML}{ffb55a}
\definecolor{myyellow}{HTML}{ffee65}
\definecolor{mypurple2}{HTML}{beb9db}
\definecolor{mypink}{HTML}{fdcce5}
\definecolor{mycyan}{HTML}{8bd3c7}

\definecolor{myblue2}{HTML}{115f9a}
\definecolor{myred2}{HTML}{c23728}

\definecolor{EOCH-color}{HTML}{ea6a4f}
\definecolor{EB-color}{HTML}{e7991b}
\definecolor{GTOM-color}{HTML}{375eac}
\definecolor{BDRC-color}{HTML}{53a559}
\definecolor{ER-color}{HTML}{126a62}
\definecolor{EP-color}{HTML}{3a8b9e}
\definecolor{EK-color}{HTML}{aa4f32}

\newtheorem{theorem}{Theorem}
\newtheorem{lemma}[theorem]{Lemma}

\title{Effective Edge Centrality via Neighborhood-based Optimization}
\author {
    Renchi Yang
}
\affiliations{
    Department of Computer Science\\
    Hong Kong Baptist University\\
    renchi@hkbu.edu.hk
}

\usepackage{bibentry}

\begin{document}

\maketitle

\begin{abstract}
Given a network $\G$, {\em edge centrality} is a metric used to evaluate the importance of edges in $\G$, which is a key concept in analyzing networks and finds vast applications involving edge ranking. In spite of a wealth of research on devising edge centrality measures, they incur either prohibitively high computation costs or varied deficiencies that lead to sub-optimal ranking quality. 

To overcome their limitations, this paper proposes \algo{}, a new centrality measure for edge ranking that is formulated based on neighborhood-based optimization objectives. We provide in-depth theoretical analyses to unveil the mathematical definitions and intuitive interpretations of the proposed \algo{} measure from diverse aspects. Based thereon, we present three linear-complexity algorithms for \algo{} estimation with non-trivial theoretical accuracy guarantees for centrality values.
Extensive experiments comparing \algo{} against six existing edge centrality metrics in graph analytics tasks on real networks showcase that \algo{} offers superior practical effectiveness while offering high computation efficiency.
\end{abstract}

%

\section{Introduction}
{\em Edge centrality} is a fundamental tool in network analysis for assessing the importance of edges within a network $\G$, which empowers us to rank edges in terms of both network topology and dynamics \cite{white2003algorithms}. 
In the past decades, edge centrality has seen widespread use in a variety of applications, such as identifying strong ties in social networks \cite{ding2011relation}, protection of infrastructure networks \cite{bienstock2014chance}, identification of failure locations in materials \cite{pournajar2022edge},  community detection \cite{fortunato2004method}, transportation \cite{jana2023edge,crucitti2006centrality}, and many others \cite{mitchell2019effectiveness,yoon2006algorithm,cuzzocrea2012edge,wang2023efficient,lai2024efficient}.



In the literature, extensive efforts have been devoted towards designing effective centrality metrics for edge ranking \cite{brohl2019centrality,kucharczuk2022pagerank}. As reviewed in Section \ref{sec:preliminary}, the majority of them can be classified into three categories: {\em ratio-based}, {\em recursive-based}, and {\em leave-one-out} centrality measures, as per their definitional styles. 
The ratio-based scheme conceives of edge importance as the fraction of sub-structures in the input network $\G$ embodying the edge. Two representatives include the well-known {\em edge betweenness} \cite{girvan2002community} and {\em effective resistance} \cite{spielman2008graph}, both of which entail tremendous computation overhead due to the enumeration of the shortest paths between all possible node pairs and minimum spanning trees in $\G$.
The leave-one-out centrality indices also suffer from severe efficiency issues, as they require calculating expensive network metrics by excluding each edge, causing a cubic time complexity.
Moreover, these centrality metrics fail to account for directions of edges, and thus, produce compromised ranking effectiveness on directed networks.

\begin{figure}[!t]
\centering
\includegraphics[width=0.8\columnwidth]{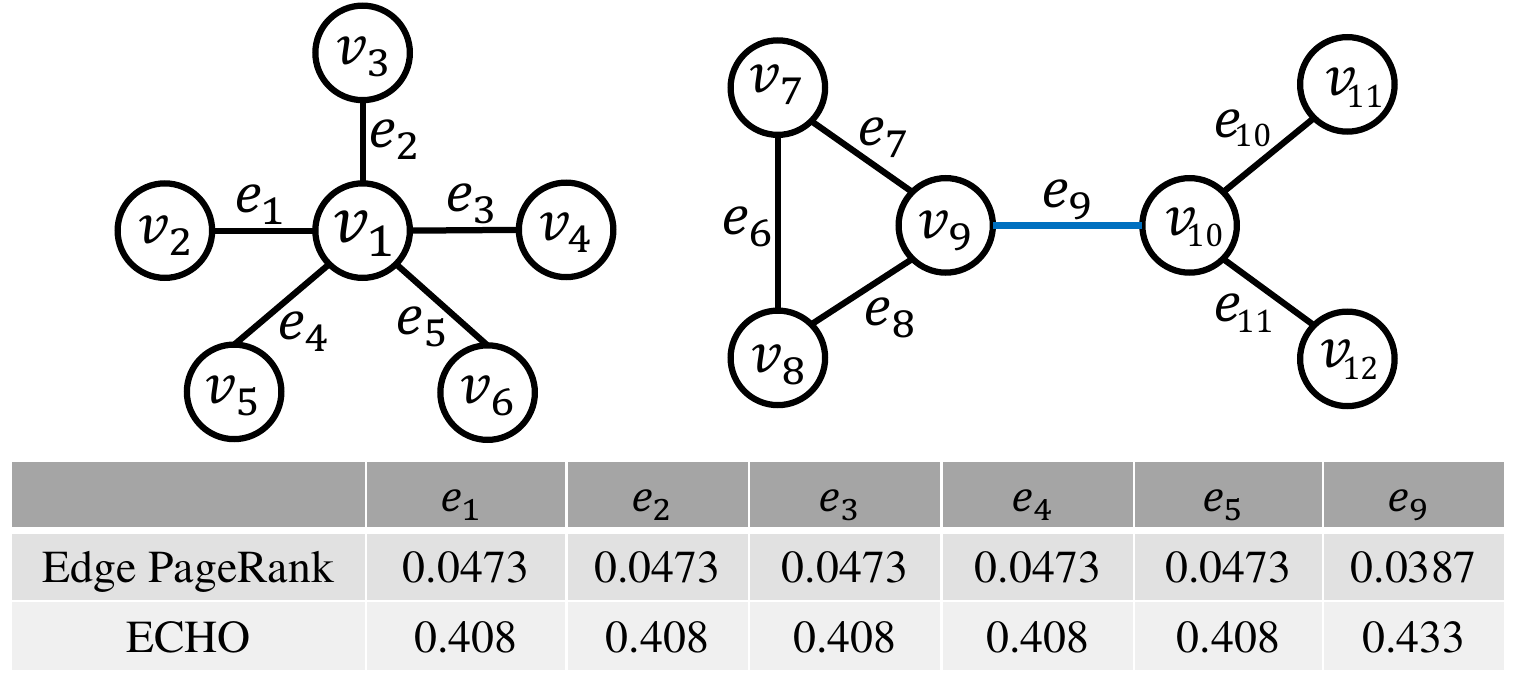}
\vspace{-3mm}
\caption{An example network $\G$}\label{fig:toy}
\vspace{-6mm}
\end{figure}

To mitigate the foregoing issues, recent works extend prominent recursive-based centrality measures for nodes (e.g., PageRank \cite{page1998pagerank}) to their edge counterparts (e.g., edge PageRank), which enable us to compute the centralities of all edges in $\G$ with linear asymptotic complexity and cope with edge directions.
Unfortunately, as pinpointed in Section \ref{sec:preliminary}, these metrics have inherent drawbacks as they quantify the edge importance merely using the strength of directed topological connections to one of its endpoints.
As a consequence, such measures strongly rely on the assumption that $\G$ is connected, as the amount and strength of connections of nodes in various components in disconnected networks are incomparable.
For instance, on the disconnected network with two components illustrated in Fig. \ref{fig:toy}, edge $e_9$ is intuitively more important to $\G$ than $e_1$-$e_5$ as the deletion of it renders nodes $v_7$-$v_9$ disconnected from nodes $v_{12}$-$v_{12}$, significantly altering the structure of $\G$. However, if we adopt edge PageRank (EP) for edge ranking, the EP of $e_9$ (i.e., $0.0387$) is much smaller than those of edges $e_1$-$e_5$ (i.e., $0.0473$), indicating that edges $e_1$-$e_5$ are more valuable to $\G$ than $e_9$ and contradicting the above intuition.
In summary, existing centrality measures for edges are either computationally demanding 
or suffer from limited ranking efficacy.

Furthermore, despite considerable research \cite{kucharczuk2022pagerank,murai2019sensitivity,bao2022benchmark} invested in theoretical analysis of these edge centrality metrics,
to our knowledge, none of the extant studies have explored their practical effectiveness in downstream graph analytics tasks. Therefore, it still remains unclear which edge centrality metrics should be adopted in real applications.

In this paper, we propose a new centrality measure \algo (\underline{E}dge \underline{C}entrality via neig\underline{H}borhood-based \underline{O}ptimization) for edge ranking in undirected or directed networks.
More specifically, \algo is formulated based on two neighborhood-based optimization objectives, where the former incorporates out-degrees of both endpoints of edges into the centrality values, while the latter forces the centrality scores of adjacent edges close. Such optimization goals require no assumptions on the network structure, i.e., connected or disconnected, and hence, enjoy better ranking effectiveness.
For instance, if we revisit the example in Fig. \ref{fig:toy}, the issue of edge PageRank is now resolved when \algo is utilized ($0.433>0.408$).
Theoretically, to facilitate the intuitive understanding of \algo,
thorough theoretical discussions of \algo from three perspectives are provided.
Building on these interpretations, we further present three approximate algorithms for \algo computation that run in time linear to the size of $\G$, followed by theoretical analyses in terms of their approximation accuracy guarantees, time complexities, or convergence.
Empirically, we demonstrate the superiority of \algo over existing edge centrality measures in practical effectiveness through extensive experiments on six real datasets in three popular network tasks involving graph clustering, network embedding, and graph neural networks.





\section{Preliminaries}\label{sec:preliminary}

\subsection{Notations and Terminology}
Throughout this paper, sets are symbolized by calligraphic letters, e.g., $\V$, and $|\V|$ is used to denote the cardinality of the set $\V$. Matrices (resp. vectors) are represented as bold uppercase (resp. lowercase) letters, e.g., $\AM$ (resp. $\xm$). We use superscript $\AM^{\top}$ to represent the transpose of matrix $\AM$. $\AM[i,j]$ denotes the $(i,j)$-th entry in matrix $\AM$.

Let $\G=(\V,\EDG)$ be a network (a.k.a. graph) with a node set $\V$ containing $n$ nodes and an edge set $\EDG$ consisting of $m$ edges.
For each node $v\in \V$, we denote by $\N^{-}(v)$ (resp. $\N^{+}(v)$) the set of incoming (resp. outgoing) neighbors of $v$. Particularly, $\N^{-}(v)=\N^{+}(v)$ when $\G$ is undirected. We use the diagonal matrix $\DM$ to represent the out-degree matrix of $\G$, where each diagonal element $\DM[v,v]=|\N^{+}(v)|$. The adjacency matrix of $\G$ is symbolized by $\AM$, in which $\AM[u,v]=1$ if $(u,v)\in \EDG$, otherwise $0$.
The incidence matrix of $\G$ is denoted by $\EM\in \mathbb{R}^{n\times m}$, wherein $\EM[v_i,e_i]=1$ if node $v_i$ is {\em incident} with edge $e_i$, and $\EM[v_i,e_i]=0$ otherwise. 

\subsection{Existing Edge Centrality Measures}\label{sec:ECM}

\begin{table*}
\centering
\renewcommand{\arraystretch}{0.7}
\caption{Existing Edge Centralities}\label{tbl:EC}
\vspace{-3mm}
\begin{footnotesize}
\begin{tabular}{|l|c|c|c|}
\hline
{\bf Centrality} & {\bf Edge Direction }   & {\bf Complexity}  \\ \hline
Edge Betweenness (EB) \cite{girvan2002community} & \Checkmark    &  $O(n^2+mn)$ \\
$\kappa$-Path Edge Centrality (KPC) \cite{de2012novel} &  \Checkmark   &  $O(m+n^2\kappa)$ \\
Effective Resistance (ER) \cite{spielman2008graph} & \XSolidBrush   &  $O(mn^\frac{3}{2})$ \\
BDRC \cite{yi2018biharmonic} & \XSolidBrush  &  $O(m+n^3)$ \\ \hline
Edge PageRank (EP) \cite{chapela2015intentional} & \Checkmark   & $O(m\log{n})$  \\ 
Edge Katz (EK) \cite{chapela2015intentional} & \Checkmark   &  $O(m\log{n})$ \\
Eigenedge (EE) \cite{huang2019eigenedge} &  \Checkmark    &  $O(m\log{n})$ \\ \hline
Information Centrality (IC) \cite{fortunato2004method} &  \Checkmark   & $O(mn^2\log{n})$  \\
$\theta$-Kirchhoff Edge Centrality ($\theta$-KEC) \cite{li2018kirchhoff} &  \XSolidBrush    & $O(mn^3)$  \\
Kemeny Edge Centrality (KEC) \cite{altafini2023edge}  & \XSolidBrush   &  $O(m^2n)$ \\ \hline
GTOM  \cite{yip2007gene} &  \Checkmark   & $O(m\cdot \max_{v\in V}{|\N^{+}(v)|})$  \\
Current-Flow Centrality (CFC) \cite{brandes2005centrality}  & \Checkmark & $O(mn\log{n})$  \\
\hline
Our \algo  & \Checkmark  &   $O(m\log{m})$ \\
\hline
\end{tabular}
\end{footnotesize}
\vspace{-1mm}
\end{table*}

In what follows, we categorize existing edge centrality metrics 
into four groups and review each of them.

\stitle{Ratio-based Edge Centralities} At a high level, this category of centralities utilizes the fraction of paths or spanning trees containing target edge $e=(u,v)$ to quantify the importance of $e$ in the network $\G$. Amid them, {\em edge betweenness} (EB) \cite{girvan2002community} is one of the most classic measures, which is formally defined by $C_{\text{EB}}(e) = \sum_{s\neq t\in V}{\frac{\delta_{s,t}(e)}{\delta_{s,t}}}$,
where $\delta_{s,t}$ signifies the number of shortest paths from $s$ to $t$ and $\delta_{s,t}(e)$ is the number of such paths passing through edge $e$. However, EB falls short of considering arbitrary paths that embody rich topological information \cite{stephenson1989rethinking}. As a remedy, in \cite{de2012novel}, the authors extend the $\kappa$-path centrality \cite{alahakoon2011k} for nodes to edges, dubbed as {\em $\kappa$-path edge centrality} (KPC). More precisely, KPC $C_{\text{KPC}}(e)$ is the sum of the factions $\frac{\delta_{s}(e)}{\delta_{s}}$ of length-$\kappa$ paths originating from every node $s\in V$ traversing edge $e$: $C_{\text{KPC}}(e) = \sum_{s\in V}{\frac{\delta_{s}((u,v))}{\delta_{s}}}$.


In lieu of using paths as in EB and KPC, {\em effective resistance} (ER) \cite{lovasz1993random,yang2023efficient} (a.k.a. {\em spanning edge betweenness/centrality} \cite{teixeira2013spanning,zhang2023efficient} or {\em resistance distance}) relies on {\em } {\em minimum spanning trees} (MSTs). Formally, the ER of edge $e=(u,v)$ is formulated as $C_{\text{ER}}(e) = \frac{\tau_G(e)}{\tau_G}$,
where $\tau_G$ is the number of distinct MSTs from $\G$ and $\tau_G(e)$ represents the number of distinct MSTs from $\G$ where edge $e$ occurs. According to \cite{spielman2008graph}, the above mathematical definition can be equivalently rewritten as: $C_{\text{ER}}(e) = \LM^{\dagger}[u,u]+\LM^{\dagger}[v,v]-2\LM^{\dagger}[u,v]$,
where $\LM^{\dagger}$ denotes the Moore-Penrose pseudo-inverse of the Laplacian (i.e., $\LM=\DM-\AM$) of $\G$. \cite{yi2018biharmonic} pinpoints that ER fails to distinguish any pair of cut edges, and then based on \cite{lipman2010biharmonic} proposes a variant of ER, i.e., {\em biharmonic distance related centrality} (BDRC) in $C_{\text{BDRC}}(e) =  {\LM^{\dagger}}^2[u,u]+{\LM^{\dagger}}^2[v,v]-2{\LM^{\dagger}}^2[u,v]$ as a remedy.


\stitle{Recursive-based Edge Centralities} Another line of research resorts to a recursive definition. That is, the influence of an edge $e=(u,v)$ is proportional to the sum of the influence scores of other edges incoming to its starting point $u$. In the spirit of such an idea, \cite{chapela2015intentional} extends prominent node-wise centrality measure PageRank \cite{page1998pagerank} to {\em edge PageRank} (EP) in Eq. \eqref{eq:EP}, where the EP of edge $e=(u,v)$ is contributed by the EP values of all edges incident to the start node $u$ and its self-importance. The parameter $\alpha\in (0,1)$ is used to control the importance of the influence from such incoming edges.
\begin{scriptsize}
\begin{equation}\label{eq:EP}
C_{\text{EP}}(e) = \frac{1}{|\N^{+}(u)|}\left( \alpha\ \sum_{x\in \N^{-}(u)}{C_{\text{EP}}((x,u))}+ 1\right)
\end{equation}
\end{scriptsize}


Akin to EP, \cite{chapela2015intentional} further extends {\em Katz index} \cite{katz1953new} to Edge Katz (EK) as follows: $C_{\text{EK}}(e) = \alpha \sum_{x\in \N^{-}(u)}{C_{\text{EK}}((x,u))}+ 1$,
\cite{huang2019eigenedge} extends {\em eigencentrality} for nodes \cite{bonacich1972factoring} to Eigenedge (EE), which can be regarded as a variant of EK.



\stitle{Leave-One-Out Edge Centralities} In this category, the importance of an edge $e=(u,v)$ is evaluated by the change after excluding $e$ from $\G$ in terms of a network metric. For example, the {\em information centrality} (IC) \cite{fortunato2004method} of $e$ is defined as the relative loss in the {\em network efficiency} obtained by removing $e=(u,v)$ from $\G$: 
$C_{\text{IC}}(e) = \frac{\epsilon(G)-\epsilon(G\setminus e)}{\epsilon(G)}$,
where $\epsilon(G)$ connotes the average {\em network efficiency} (i.e., the reciprocal of the distance of two nodes) \cite{latora2001efficient} of $\G$.
In analogy to IC, $\theta$-Kirchhoff edge centrality ($\theta$-KEC) of edge $e$ is defined as the Kirchhoff index of the network obtained from $\G$ by {\em deleting} $e$ \cite{li2018kirchhoff}, where the Kirchhoff index of a network is the sum of effective resistances of all node pairs.
Very recently, \cite{altafini2023edge} propose {\em Kemeny-based Edge Centrality} (KEC): $C_{\text{KEC}}(e) = K(G\setminus {e}) - K(G),$
where $K(G)$ represents the Kemeny's constant \cite{kemeny1960finite} of $\G$, i.e., the expected number of time steps needed from a starting node $u$ to a random destination node $v$ sampled from the stationary distribution of the Markov chain induced by the transition matrix of $\G$.






\stitle{Others} The {\em generalized topological overlap matrix} (GTOM) \cite{yip2007gene} of $e$ simply measures the amount of common direct successors that the start and end nodes of $e$ share, i.e., $C_{\text{GTOM}}(e) = \frac{|\N^{+}(u)\cap \N^{+}(v)|+1}{\min\{|\N^{+}(u)|,|\N^{+}(v)|\}}$.
The {\em current-flow centrality} (CFC) \cite{brandes2005centrality} leverages the number of flows between any two nodes $s,t\in V$ that pass through $e$ as its centrality.

Table \ref{tbl:EC} summarizes the properties and computational complexities of the above 12 edge centrality metrics. Note that only three existing edge centrality measures, i.e., EP, EK, and EE, can be applied to directed edges and achieve a linear asymptotic performance.
By \cite{chapela2015intentional}, $C_{\text{EP}}(e)$ and $C_{\text{EK}}(e)$ are essentially reweighted PageRank $PR(u)$ and Katz index $KA(u)$ of node $u$, respectively, i.e., $C_{\text{EP}}(e)= \frac{PR(u)}{|\N^{+}(u)|}$ and $\ C_{\text{EK}}(e)= KA(u)$.
Such a definition overlooks the topological influences from other edges on $e$ via node $v$, leading to undermined edge ranking quality. EE has the same flaw as it is a special case of EK. In addition, EK produces extremely high centrality values, rendering it hard to distinguish vital edges from trivial ones. Most importantly, all these metrics rely on the strength of connections between nodes and require $G$ to be connected, and hence, fail on disconnected networks, as exemplified in Fig. \ref{fig:toy}.

\section{The \algo Measure}
In this section, we first elaborate on the design of \algo, followed by the interpretations of \algo, algorithmic details for computing \algo, and related theoretical analysis.

\subsection{Optimization Objective}
Distinct from existing measures, we formulate the definition of \algo through an optimization problem with two goals.
More concretely, we represent by $\zm\in \mathbb{R}^m$ the \algo vector, wherein $\zm[e_i]$ signifies the \algo value of edge $e_i$.
In mathematical terms, \algo vector $\zm$ can be obtained by optimizing the following neighborhood-related objectives:
\begin{footnotesize}
\begin{equation}\label{eq:obj}
\min_{\zm\in \mathbb{R}^m}{\mathcal{L}=(1-\alpha) \|\zm-\xm\|^2 + \alpha \sum_{e_i,e_j\in \EDG, i<j}\frac{1}{2}{\sum_{v\in e_i\cap e_j}\frac{\left(\zm_i-\zm_j\right)^2}{|\N^{+}(v)|}}},
\end{equation}
\end{footnotesize}
where $\forall{e_i\in \EDG}$
\begin{footnotesize}
\begin{equation}\label{eq:xm}
\xm[e_i] = \frac{1}{\sqrt{|\N^{+}(u_i)|+|\N^{+}(v_i)|}},
\end{equation}
\end{footnotesize}
and $\alpha\in (0,1)$ is a weight balancing two optimization terms.

Intuitively, if the endpoints $u_i,v_i$ of edge $e_i$ are scarcely connected to others, deleting $e_i$ from $\G$ is more likely to disconnect nodes $u_i$, $v_i$ and the subgraphs containing them, engendering a considerable impact on the connectivity and structure of the entire graph. In other words, an edge $e_i$ is less (resp. more) important to $\G$ if its endpoints are connected to a multitude of (resp. a small number of) nodes/edges, i.e., $|\N^{+}(u_i)|+|\N^{+}(v_i)|$ is large (resp. small). The first goal $\min_{\zm\in \mathbb{R}^m}\|\zm-\xm\|^2$ in Eq. \eqref{eq:obj} enforces $\zm[e_i]$ close to $\xm[e_i] = \frac{1}{\sqrt{|\N^{+}(u_i)|+|\N^{+}(v_i)|}}$ in Eq. \eqref{eq:xm}, which assigns a high (resp. low) importance value to $e_i$ with a paucity of (resp. massive) neighbors.

On the other hand, the objective of \algo in Eq. \eqref{eq:obj} additionally requires the centrality values of edges with common endpoints to be close. For example, given two edges $e_i,e_j$ incident from the same node $v$, the difference between their \algo values $\left(\zm_i-\zm_j\right)^2$ should be minimized. Considering that each edge has two endpoints and there are $|\N^{+}(v)|$ edges incident from $v$ in total, the average difference between two edges $e_i,e_j$ can be quantified as $\frac{1}{2}{\sum_{v\in e_i\cap e_j}\frac{\left(\zm_i-\zm_j\right)^2}{|\N^{+}(v)|}}$, leading to our second optimization goal in  Eq. \eqref{eq:obj}.

\begin{theorem}\label{lem:C-def}
The optimal solution to Eq. \eqref{eq:obj} is expressed by 
\begin{footnotesize}
\begin{gather}
\zm = (1-\alpha)\cdot\left(\IM - {\alpha}\cdot \frac{1}{2}{\EM}^{\top}\DM^{-1}\EM\right)^{-1}\xm.\label{eq:zm_inverse}
\end{gather}
\end{footnotesize}
\end{theorem}

Our theoretical analysis in Theorem \ref{lem:C-def} states the closed-form solution of \algo vector $\zm$ to Eq. \eqref{eq:obj}. In the succeeding section, we transform it into its equivalent mathematical definitions and interpret the definitions from three perspectives.


\eat{
Akin to the recursive-based edge centrality measures introduced in Section \ref{sec:ECM}, we define the \algo $C_{\text{\algoabbr}}(e)$ of edge $e=(u,v)$ as the weighted sum of ERWR scores of $e$ w.r.t. the all possible source edges $e_i$ in $\G$:
\begin{small}
\begin{equation}\label{eq:EdgeRWR}
C_{\text{\algoabbr}}(e)=\sum_{e_i=(u_i,v_i)\in \EDG}{\frac{ r(e_i,e)}{\sqrt{|\N^{+}(u_i)|+|\N^{+}(v_i)|}}},
\end{equation}
\end{small}
where each source edge $e_i$ is weighted by $\frac{1}{\sqrt{|\N^{+}(u_i)|+|\N^{+}(v_i)|}}$. This weight is to reflect the importance of $e_i$'s influence (ERWR score) on $e$ from the perspective of the whole $\G$.
}

\subsection{Interpretations of \algo}

\subsubsection{\bf Spectral Graph Signal Filtering}\label{sec:inter-spectral}
On the basis of Theorem \ref{lem:C-def}, we can derive the following lemma:
\begin{lemma}\label{lem:C-def-eig}
The optimal solution to Eq. \eqref{eq:obj} is $\zm = \UM\frac{1}{\IM-\alpha^2\boldsymbol{\Sigma}^2}\UM^{\top}\xm$,
where matrix $\UM$ and diagonal matrix $\boldsymbol{\Sigma}$ contain the left singular vectors and singular values of $\frac{1}{\sqrt{2}}\EM\DM^{-\frac{1}{2}}$, respectively.
\end{lemma}

To interpret this eigendecomposition-based definition of \algo, we first regard $\frac{1}{\sqrt{2}}\EM\DM^{-\frac{1}{2}}\in \mathbb{R}^{m\times n}$ as a data matrix containing $m$ data points with $n$ dimensions. Thus, $\frac{1}{2}{\EM}^{\top}\DM^{-1}\EM$ can represent the adjacency matrix of the complete graph $\widetilde{\G}$ with $m$ data points as nodes, in which each edge $(e_i,e_j)$ is associated with a weight $\frac{1}{2}({\EM}^{\top}\DM^{-1}\EM)[e_i,e_j]$. The graph Laplacian of $\widetilde{\G}$ can be derived by $\widetilde{\LM}=\IM-\frac{1}{2}{\EM}^{\top}\DM^{-1}\EM$. Let its eigendecomposition be $\widetilde{\LM} = \widetilde{\UM}\widetilde{\boldsymbol{\Lambda}}\widetilde{\UM}^{\top}$, where $\widetilde{\UM}$ contains the eigenvectors and the diagonal matrix $\widetilde{\boldsymbol{\Lambda}}$ consists of the eigenvalues of $\widetilde{\LM}$. Recall that in graph signal processing~\cite{ortega2018graph}, the graph Fourier transform of a signal $\sm \in \mathbb{R}^{m}$ is defined as $\widetilde{\sm}=\widetilde{\UM}^{\top}\sm$ and the inverse transformation is $\sm = \widetilde{\UM}\widetilde{\sm}$. The transform enables the formulation of operations such as filtering in the spectral domain.
The filtering operation on signals $\sm$ with a filter $g(\cdot)$ is defined as
\begin{equation}\label{eq:s-filter}
\sm^{\circ} = g(\LM)\sm = g(\widetilde{\UM}\boldsymbol{\Lambda}\widetilde{\UM}^{\top})\sm = \widetilde{\UM} g(\boldsymbol{\Lambda})\widetilde{\UM}^{\top}\sm.
\end{equation}
\begin{lemma}\label{lem:U-Lambda-Sigma}
Let $\UM\boldsymbol{\Sigma}\VM^{\top}$ be the singular value decomposition (SVD) of $\frac{1}{\sqrt{2}}{\EM}^{\top}\DM^{-\frac{1}{2}}$. Then, $\widetilde{\UM}=\UM$ and $\widetilde{\boldsymbol{\Lambda}}=\IM-\boldsymbol{\Sigma}^2$.
\end{lemma}
Using Lemma \ref{lem:U-Lambda-Sigma}, we can further transform the filtered signal $\sm^{\circ}$ in Eq. \eqref{eq:s-filter} into its equivalent form
\begin{equation*}
\sm^{\circ} = \UM g(\IM-\boldsymbol{\Sigma}^2)\UM^{\top}\sm.
\end{equation*}
By Lemma \ref{lem:C-def-eig}, if we let $\sm=\xm$ and $\sm^{\circ}=\zm$, the above equation implies that \algo vector $\zm$ is essentially the filtered signal of $\xm$ with filtering function $g(x)=\frac{1}{1-\alpha^2+\alpha^2 x}$ on graph $\widetilde{\G}$.

\subsubsection{\bf Markov Random Walks}\label{sec:inter-rw}
Eq. \eqref{eq:zm_inverse} in Theorem \ref{lem:C-def} can be further converted into its equivalent form in Eq. \eqref{eq:zm} in Lemma \ref{lem:C-def-sum} using the Neuman series.
\begin{lemma}\label{lem:C-def-sum}
The optimal solution to Eq. \eqref{eq:obj} is represented as 
\begin{footnotesize}
\begin{equation}\label{eq:zm}
\textstyle \zm = \BM\xm\ \text{where}\ \BM=\sum_{\ell=0}^{\infty}{(1-\alpha)\alpha^{\ell}{\left(\frac{1}{2}{\EM}^{\top}\DM^{-1}\EM\right)}^{\ell}}. 
\end{equation}
\end{footnotesize}
\end{lemma}

Next, we expound on the definition of \algo in Eq. \eqref{eq:zm} using Markov random walks. First, we define the {\em edge-wise random walk} (ERW) over $\G$. Given network $\G$, a jumping probability $\alpha$, a source edge $e_s=(u_s,v_s)$ and a target edge $e_t=(u_t,v_t)$, an ERW on $\G$ originating from $e_s$ proceeds as follows:
at each step, the walk either (i) terminates at the current edge $e_i=(u_i,v_i)$ with a probability of $1-\alpha$; or (ii) with the remaining $\alpha$ probability, jumps to node $u_j$ that is incident with $e_i$ ($u_j=v_i$ when $\G$ is directed and $u_j=u_i$ or $v_i$ with an equal likelihood when $\G$ is undirected), and subsequently navigates to one of the out-going edges $(u_j,v_j)$ of the current node $u_j$ (i.e., $u_i$ or $v_i$) according to the probability of $\frac{1}{|\N^{+}(u_j)|}$.

We refer to the probability of the above walk from $e_s$ stopping at $e_t$ in the end as the ERW score of $e_t$ w.r.t. $e_s$ and denote it by $r(e_s,e_t)$. Intuitively, $r(e_s,e_t)$ quantifies the strength of {\em multi-hop} connections from $e_s$ to $e_t$ with consideration of the edge directions, which can be interpreted as the total direct and indirect influences of $e_s$ on $e_t$. By its definition, it can be proved that $r(e_s,e_t)=\BM[e_i,e_j]$ and using Lemma \ref{lem:C-def-sum} yields the following lemma:
\begin{lemma}\label{lem:ERWR}
Let $r(e_i,e_j)$ be the ERW score of edges $e_i,e_j$.
\begin{scriptsize}
\begin{align*}
\zm[e_i]& =\sum_{e_j\in \EDG}{\frac{r(e_i,e_j)}{\sqrt{|\N^{+}(u_j)|+|\N^{+}(v_j)|}}}=\sum_{e_j\in \EDG}{\frac{r(e_j,e_i)}{\sqrt{|\N^{+}(u_j)|+|\N^{+}(v_j)|}}},
\end{align*}
\end{scriptsize}
which ranges from ${1}/{\underset{{e_i\in \EDG}}{\max}{\sqrt{|\N^{+}(u_i)|+|\N^{+}(v_i)|}}}$ to ${1}/{\underset{{e_i\in \EDG}}{\min}{\sqrt{|\N^{+}(u_i)|+|\N^{+}(v_i)|}}}$.
\end{lemma}
Lemma \ref{lem:ERWR} indicates that the \algo $\zm[e_i]$ of edge $e_i$, is the sum of ERW scores of all edges in $\G$ when $e_i$ is source or target, i.e., the total influence from all edges to $e_i$ or from $e_i$ to all edges.

\subsubsection{\bf Element-wise Scaled Dominant Eigenvector}\label{sec:inter-eig}
According to Eq. \eqref{eq:zm_inverse}, we note that $\zm = (1-\alpha)\xm + \frac{\alpha}{2}{\EM}^{\top}\DM^{-1}\EM\zm$.
If we define $\boldsymbol{\Delta}$ as an $m$ by $m$ diagonal matrix wherein each $e_i$-th diagonal entry is $\boldsymbol{\Delta}[e_i,e_i]={m\cdot\xm[e_i]}$. Due to the symmetric property of ${\EM}^{\top}\DM^{-1}\EM$, after multiplying $\boldsymbol{\Delta}^{-1}$, we can rewrite the above equation as follows:
\begin{footnotesize}
\begin{equation}\label{eq:ym}
\textstyle \ym = (1-\alpha)\cdot \frac{\mathbf{1}}{m} + \frac{\alpha}{2}{\EM}^{\top}\DM^{-1}\EM \ym,
\end{equation}
\end{footnotesize}
where $\ym = \boldsymbol{\Delta}^{-1}\zm$. 
By Lemma \ref{lem:ERWR}, $\|\ym\|_1=\frac{1}{m}\sum_{e_i,e_j\in \EDG}{r(e_i,e_j)}= \frac{1}{m}\sum_{e_i\in \EDG}\|\BM[e_i]\|_1$.
\begin{lemma}\label{lem:2property}
Given any $\alpha\in (0,1)$ and integer $\ell\ge 1$, $\BM$ and ${\left(\frac{1}{2}{\EM}^{\top}\DM^{-1}\EM\right)}^{\ell}$ are semi-positive bistochastic matrices.
\end{lemma}

Note that by Lemma \ref{lem:2property}, $\BM$ is a semi-positive bistochastic matrix satisfying $\|\BM[e_i]\|_1=1\ \forall{e_i\in \EDG}$. As such, Eq. \eqref{eq:ym} can be transformed into
\begin{footnotesize}
\begin{equation}\label{eq:ym-H}
\ym = \HM \ym,\ \text{where}\ \HM=(1-\alpha)\cdot \frac{\mathbf{1}\cdot \mathbf{1}^{\top}}{m} + \frac{\alpha}{2}{\EM}^{\top}\DM^{-1}\EM,
\end{equation}
\end{footnotesize}
implying that $\ym$ is an eigenvector of matrix $\HM$ whose corresponding eigenvalue is $1$. 
Furthermore, notice that both $\frac{\mathbf{1}\cdot \mathbf{1}^{\top}}{m}$ and $\frac{1}{2}{\EM}^{\top}\DM^{-1}\EM$ (Lemma \ref{lem:2property}) are bistochastic matrices. Hence, $\HM$ is a positive bistochastic matrix, which is also irreducible. By Perron–Frobenius theorem~\cite{horn2012matrix}, $\HM$'s largest eigenvalue (in magnitude) is then $1$. As a consequence, $\ym$ is the {\em dominant eigenvector} of $\HM$ in Eq. \eqref{eq:ym-H}.
Given that $\zm=\boldsymbol{\Delta}\ym$, $\zm$ is therefore an element-wise scaled version of the dominant eigenvector of $\HM$.


\begin{algorithm}[!t]
\caption{Simple ISM}\label{alg:algo}
\KwIn{$G=(V,E)$, weight $\alpha$ and error threshold $\epsilon$}
\KwOut{$\zm^{\prime}[e]\ \forall{e\in \EDG}$}
Initialize a column vector $\xm\in \mathbb{R}^{m}$\;
\lFor{$e_i\in \EDG$}{
Initialize $\xm[e_i]$ as in Eq. \eqref{eq:xm}
}
$\zm^{\prime} \gets \xm$; $t\gets \lceil\log_{\alpha}{\epsilon}-1\rceil$\;
\lFor{$i\gets 1$ to $t$}{
Compute $\zm^{\prime}$ according to Eq. \eqref{eq:compute-z}
}
\Return{$\zm^{\prime}[e]\ \forall{e\in \EDG}$}\;
\end{algorithm}

\begin{algorithm}[!t]
\caption{Adaptive ISM}\label{alg:algo2}
\KwIn{$G=(V,E)$, weight $\alpha$ and error threshold $\epsilon$}
\KwOut{$\zm^{\prime}[e]\ \forall{e\in \EDG}$}
{\nonl{Lines 1-2 are the same as Lines 1-2 in Algo. \ref{alg:algo}}\;
\setcounter{AlgoLine}{2}}
$\thetam \gets \xm;\ \zm^{\prime} \gets (1-\alpha)\cdot \thetam$\;
\While{$\exists e_i\in \EDG$ such that $\thetam[e_i]> \epsilon$}{
Update $\thetam$ according to Eq. \eqref{eq:update-r}\;
Increase $\zm^{\prime}$ by $(1-\alpha)\cdot \thetam$\;
}
\Return{$\zm^{\prime}[e]\ \forall{e\in \EDG}$}\;
\end{algorithm}

\section{Algorithms}\label{sec:algos}
Since the definition of \algo introduced in Section \ref{sec:inter-spectral} demands expensive eigendecomposition ($O(mn)$ time and space), we resort to leveraging Eq. \eqref{eq:zm} and Eq. \eqref{eq:ym-H} for the estimation of the \algo vector in time linear to the size of $\G$, as delineated in Sections~\ref{sec:IAM} and \ref{sec:DEV}, respectively.

\subsection{Iterative Summation Methods}\label{sec:IAM}
Recall that in Eq. \eqref{eq:zm}, $\zm$ involves summing up an infinite series of matrix-vector multiplications, making exact computation infeasible. A simple and straightforward way of approximating $\zm$ is to calculate its truncated version with a maximum number of iterations. 

\stitle{Simple ISM}
In Algo.~\ref{alg:algo}, we display the pseudo-code of this approach, which attains an absolute error $\epsilon$ in the estimated \algo of each edge in $\G$ with a fixed number $t$ of iterations.
Algo.~\ref{alg:algo} begins by taking as input the network $\G$, weight $\alpha$, and error threshold $\epsilon$. Afterwards, Algo.~\ref{alg:algo} initializes a length-$m$ column vector $\xm$ according to Eq. \eqref{eq:xm} (Lines 1-2). After initializing $\zm^{\prime}$ as $\xm$ at Line 3, Algo.~\ref{alg:algo} starts an iterative process to update $\zm^{\prime}$ (Line 4). Specifically, in each iteration, we compute a new $\zm^{\prime}$ by
\begin{footnotesize}
\begin{equation}\label{eq:compute-z}
\zm^{\prime} \gets (1-\alpha)\cdot\xm + \frac{\alpha}{2}\cdot {\EM}^{\top}\DM^{-1}\cdot(\EM \zm^{\prime}).
\end{equation}
\end{footnotesize}
After repeating the above step for $t=\lceil\log_{\alpha}{\epsilon}-1\rceil$ iterations,  
Algo.~\ref{alg:algo} finally returns $\zm^{\prime}[e]\ \forall{e\in \EDG}$ as the approximate \algo values of all edges in $\G$.
Theorem \ref{lem:approx-err} establishes the approximation accuracy guarantees for Algo.~\ref{alg:algo}.
\begin{theorem}\label{lem:approx-err}
Let $\zm^{\prime}[e]\ \forall{e\in \EDG}$ be the output of Algo. \ref{alg:algo}. Then, for each edge $e\in \EDG$, the following inequality holds:
\begin{equation}\label{cc-abs-err}
0 \le \zm[e] - \zm^{\prime}[e] \le \epsilon.
\end{equation}
\end{theorem}

\stitle{Adaptive ISM} To achieve the accuracy assurance in Eq. \eqref{cc-abs-err}, Algo. \ref{alg:algo} determines the maximum number $t$ of iterations needed solely based on $\alpha$ and $\epsilon$, which is data oblivious and demands more iterations. To fill this gap, we propose to compute the truncated version of Eq. \eqref{eq:zm} in an iterative and adaptive fashion. More specifically, we maintain a residual vector $\thetam$ in the iterative process such that $\zm = \zm^{\prime} + \BM\thetam$ holds. By repeatedly converting $\thetam$ into $\zm^{\prime}$ until the values in $\thetam$ reach a certain threshold, the gap between $\zm$ and $\zm^{\prime}$ is reduced to satisfy Eq. \eqref{cc-abs-err}. Algo. \ref{alg:algo2} illustrates the pseudo-code of this adaptive approach. After initializing $\xm$ as in Algo. \ref{alg:algo} at Lines 1-2, Algo. \ref{alg:algo2} proceeds to set the residual vector $\thetam$ as $\xm$ and approximate \algo vector $\zm^{\prime}$ as $(1-\alpha)\cdot \thetam$ (Line 3). Subsequently, if there exists any edge $e_i\in \EDG$ such that $\thetam[e_i]>\epsilon$, we update $\thetam$ by
\begin{footnotesize}
\begin{equation}\label{eq:update-r}
\thetam \gets \frac{\alpha}{2}\cdot {\EM}^{\top}\DM^{-1}\cdot(\EM \thetam)
\end{equation}
\end{footnotesize}
and $\zm^{\prime}$ is then updated as $\zm^{\prime} + (1-\alpha)\cdot \thetam$ (Lines 4-6). Otherwise, Algo. \ref{alg:algo2} ceases the above procedure and returns $\zm^{\prime}$ as the output.
As stated in Theorem \ref{lem:algo2}, $\zm^{\prime}$ is an approximate version of $\zm$ with at most $\epsilon$ additive error in each element.
\begin{theorem}\label{lem:algo2}
Let $\zm^{\prime}[e]\ \forall{e\in \EDG}$ be the output of Algo. \ref{alg:algo2}. Then, for each edge $e\in \EDG$, $0 \le \zm[e] - \zm^{\prime}[e] \le \epsilon$.
\end{theorem}

\subsubsection{\bf Complexity Analysis} According to Algo. \ref{alg:algo}, the computation expenditure for approximate \algo values of all edges in the input network $\G$ lies in the $t$ iterations of matrix multiplications (Line 4). Note that $\widehat{\EM}$ and $\EM$ are sparse matrices containing $m$ non-zero entries, and we can utilize the trick in Eq. \eqref{eq:compute-z} to reorder the sparse matrix-vector multiplications for higher efficiency. In doing so, the execution of each iteration in Line 4 takes $O(m)$ time, and thus, the time complexity for $t$ iterations is $\textstyle O\left(m\log_{\frac{1}{\alpha}}{\frac{1}{\epsilon}}\right)$ in total.

Suppose that Algo. \ref{alg:algo2} stops after $t$ rounds of Lines 5-6. Then, Algo. \ref{alg:algo2} runs in $O(mt)$ time. Each iteration updates $\thetam$ via Eq. \eqref{eq:update-r}, whose initial value is $\xm$. After $t$ iterations,
$\thetam = \alpha^t\left(\frac{1}{2} {\EM}^{\top}\DM^{-1}\EM\right)^t\xm$. By Lemma \ref{lem:2property}, $\frac{1}{2} {\EM}^{\top}\DM^{-1}\EM$ is bistochastic.
Since Algo. \ref{alg:algo2} stops when all entries in $\thetam$ are not greater than $\epsilon$, we have $\textstyle \alpha^t \underset{e_i\in \EDG}{\max}\underset{{e_j\in \EDG}}{\sum}{\left(\frac{1}{2} {\EM}^{\top}\DM^{-1}\EM\right)^t[e_i,e_j]\cdot \xm[e_j]} \le \epsilon$, 
which leads to
\begin{scriptsize}
\begin{equation*}
t=\left\lceil\log_{\frac{1}{\alpha}}{\underset{e_i\in \EDG}{\max}\underset{e_j\in \EDG}{\sum}{\left(\frac{1}{2} {\EM}^{\top}\DM^{-1}\EM\right)^t[e_i,e_j]\cdot \xm[e_j]}\cdot \frac{1}{{\epsilon}}}\right\rceil \le \left\lceil\log_{\frac{1}{\alpha}}{\frac{1}{\epsilon}}\right\rceil.
\end{equation*}
\end{scriptsize}
Overall, Algo. \ref{alg:algo2} takes $\textstyle \textstyle O\left(m\log_{\frac{1}{\alpha}}{\frac{1}{\epsilon}}\right)$ time in the worst case. 



\eat{
\renchi{

The stationary distribution of an irreducible aperiodic finite Markov chain is uniform if and only if its transition matrix is doubly stochastic
\url{https://en.wikipedia.org/wiki/Doubly_stochastic_matrix}

Birkhoff–von Neumann decomposition

\url{https://openreview.net/pdf?id=IpsTSvfIB6}

\url{https://www.mdpi.com/2073-8994/12/3/369}

\url{https://nvlpubs.nist.gov/nistpubs/jres/80B/jresv80Bn4p433_A1b.pdf}

\url{https://core.ac.uk/download/pdf/82294475.pdf}

\url{https://link.springer.com/content/pdf/10.1007/BF00533407.pdf}

\url{https://www.sciencedirect.com/science/article/pii/0012365X75900126}

\url{https://arxiv.org/abs/1310.1273}

\url{http://ijesi.org/papers/Vol(3)8/Version-2/B0382011016.pdf}
}
}

\begin{algorithm}[!t]
\caption{DEV Method}\label{alg:algo3}
\KwIn{$G=(V,E)$, weight $\alpha$ and integer $t$}
\KwOut{$\zm^{\prime}[e]\ \forall{e\in \EDG}$}
Initialize $\ym$ as $\frac{\mathbf{1}}{m}$\;
\lFor{$i\gets 1$ to $t$}{
Update $\ym$ according to Eq. \eqref{eq:update-y}
}
\lFor{$e_i\in \EDG$}{
Compute $\zm^{\prime}[e_i]$ as in Eq. \eqref{eq:compute-zm-prime} 
}
\Return{$\zm^{\prime}[e]\ \forall{e\in \EDG}$}\;
\end{algorithm}

\subsection{DEV Method}\label{sec:DEV}
Distinct from the iterative summation methods in Section \ref{sec:IAM}, the DEV method relies on the definition of the \algo vector $\zm$ in Section \ref{sec:inter-eig}. More concretely, we can first compute the dominant eigenvector $\ym$ of $\HM$ in Eq. \eqref{eq:ym-H}. The resulting $\ym$ is subsequently scaled via diagonal matrix $\boldsymbol{\Delta}$, i.e., $\boldsymbol{\Delta}\ym$, to construct $\zm^{\prime}$. 
Algo. \ref{alg:algo3} presents the pseudo-code of this DEV method. Particularly, we adopt the prominent {\em power method}~\cite{horn2012matrix} for eigenvector computation. That is, it begins with an initial uniform vector $\frac{\mathbf{1}}{m}$ as $\ym$ and then computes successive $t$ iterates $\HM\ym$ (Lines 1-2). The update operation $\HM\ym$ is implemented by 
\begin{footnotesize}
\begin{equation}\label{eq:update-y}
\ym \gets (1-\alpha)\cdot \frac{\mathbf{1}}{m}\cdot(\mathbf{1}^{\top}\ym) + \frac{\alpha}{2}{\EM}^{\top}\DM^{-1}\cdot(\EM\ym),
\end{equation}
\end{footnotesize}
whose matrix-vector multiplication orders eliminate the need for materializing $\mathbf{1}\cdot\mathbf{1}^{\top}$ and $\BM$, thereby enabling a linear update cost for each iteration. Next, for each edge $e_i\in \EDG$, $\zm^{\prime}[e_i]$ is calculated by
\begin{footnotesize}
\begin{equation}\label{eq:compute-zm-prime}
\zm^{\prime}[e_i]\gets \frac{m}{\sqrt{|\N^{+}(u_j)|+|\N^{+}(v_j)|}}\cdot \ym[e_i],
\end{equation}
\end{footnotesize}
and returned as the estimation of $\zm[e_i]$ (Lines 3-4).

\subsubsection{\bf Analysis}
According to \cite{wilkinson1988algebraic}, the convergence rate of the power method is given by $|\lambda_2|/|\lambda_1|$, where $\lambda_1$ and $\lambda_2$ stand for the largest and second largest eigenvalues of $\HM$ in terms of absolute value. Given $|\lambda_1|=1$ as analyzed in Section \ref{sec:inter-eig}, the convergence rate of Algo. \ref{alg:algo3} is $|\lambda_2|$.
Note that each iteration in Line 2 involves sparse matrix multiplication in Eq. \eqref{eq:update-y}, which consumes $O(m)$ time. Hence, the total time cost of Algo. \ref{alg:algo3} is bounded by $O(mt)$.

\section{Experiments}
This section empirically studies the effectiveness of \algo and existing edge centralities for identifying influential edges in six real-world networks through three popular graph analytics tasks, i.e., graph clustering, unsupervised network embedding, and graph neural networks, as well as their computation efficiency.
Table~\ref{tab:datasets} summarizes the statistics of the datasets used in the experiments. 
More details regarding datasets, running environments, and parameter study are deferred to our supplementary material.

\begin{table}[!t]
\renewcommand{\arraystretch}{0.8}
\caption{Statistics of Datasets}
\label{tab:datasets}
\vspace{-3mm}
\centering
\begin{footnotesize}
\resizebox{\columnwidth}{!}{%
\begin{tabular}{c|c|c|c|c}
\hline
{\bf Name} & {\bf \#Nodes} & {\bf \#Edges} & {\bf \#Attributes} & {\bf \#Classes} \\ \hline
{\em Email-EU}     & 1,005 & 25,571 & - & - \\
{\em Facebook}     & 4,039 & 88,234 & - & - \\
{\em PPI}     & 3,890 & 76,584 & - & 50 \\
{\em BlogCatalog}     & 10,312 & 333,983 & - & 39 \\
{\em Cora}     & 2,708 & 10,556 & 1,433 & 7 \\
{\em Chameleon}     & 2,277 & 62,792 &  2,325 & 5 \\
\hline
\end{tabular}
}
\end{footnotesize}
\vspace{-2ex}
\end{table}

We evaluate our proposed \algo against the six best competitors, EB, ER, EP, EK, GTOM, and BDRC, out of all the edge centrality measures listed in Table \ref{tbl:EC}. We exclude KPC, KEC, CFC, and IC from comparison due to their inferior ranking efficacy as well as extremely high computation overheads. Additionally, EE can be regarded as a variant of EP or EK, and thus, is omitted. Unless specified otherwise, for \algo, we set $\alpha=0.5$ and the maximum number of iterations to $150$ (i.e., $\epsilon=10^{-45}$ in Algo. \ref{alg:algo}). 
For reproducibility, all the codes and datasets are made publicly available at 
{\url{https://github.com/HKBU-LAGAS/ECHO}}.





\subsection{Effectiveness Evaluation}
In this set of experiments, we evaluate the effectiveness of \algo and competing edge centralities as follows. After calculating the centrality values of the edges, we sort all edges of the input network $G$ in ascending order by their centrality values. Then, we remove the top-$(m\cdot \rho)$ ($\rho$ is varied from 0.1 to 0.9) edges from $G$ to create a residual network $G^{\prime}$, followed by inputting it to the classic methods, {\em spectral clustering} \cite{von2007tutorial}, node2vec \cite{grover2016node2vec}, and GCN model \cite{kipf2016semi} for node clustering, node classification, and semi-supervised node classification, respectively. 

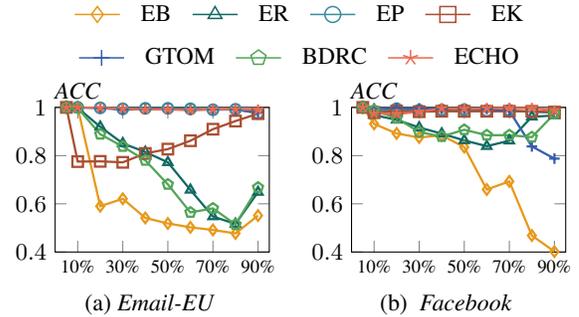
\begin{figure}[!t]
\centering
\begin{small}
\begin{tikzpicture}
    \begin{customlegend}[legend columns=4,
        legend entries={EB, ER, EP, EK},
        legend style={at={(0.45,1.35)},anchor=north,draw=none,font=\footnotesize,column sep=0.2cm}]
    \addlegendimage{line width=0.2mm,mark size=3pt,color=EB-color,mark=diamond}
    \addlegendimage{line width=0.2mm,mark size=3pt,color=ER-color,mark=triangle}
    \addlegendimage{line width=0.2mm,mark size=3pt,color=EP-color,mark=o}
    \addlegendimage{line width=0.2mm,mark size=3pt,color=EK-color,mark=square}
    \end{customlegend}
\end{tikzpicture}
\begin{tikzpicture}
    \begin{customlegend}[legend columns=3,
        legend entries={GTOM, BDRC, \algo},
        legend style={at={(0.45,0.35)},anchor=north,draw=none,font=\footnotesize,column sep=0.2cm}]
    \addlegendimage{line width=0.2mm,mark size=3pt,color=GTOM-color,mark=+}
    \addlegendimage{line width=0.2mm,mark size=3pt,color=BDRC-color,mark=pentagon}
    \addlegendimage{line width=0.2mm,mark size=3pt,color=EOCH-color,mark=star}
    \end{customlegend}
\end{tikzpicture}
\\[-\lineskip]
\vspace{-4mm}
\subfloat[{\em Email-EU}]{
\begin{tikzpicture}[scale=1,every mark/.append style={mark size=2pt}]
    \begin{axis}[
        height=\columnwidth/2.4,
        width=\columnwidth/1.9,
        ylabel={\it ACC},
        xmin=0, xmax=9.5,
        ymin=0.4, ymax=1.0,
        xtick={1,3,5,7,9},
        xticklabel style = {font=\scriptsize},
        yticklabel style = {font=\small},
        xticklabels={10\%,30\%,50\%,70\%,90\%},
        every axis y label/.style={font=\small,at={(current axis.north west)},right=3mm,above=0mm},
        legend style={fill=none,font=\small,at={(0.02,0.99)},anchor=north west,draw=none},
    ]
    \addplot[line width=0.3mm, color=EB-color,mark=diamond]  
        plot coordinates {
(0.5,	1.0000	)
(1,	1	)
(2,	0.59005	)
(3,	0.620896	)
(4,	0.541294	)
(5,	0.518408	)
(6,	0.502488	)
(7,	0.491542	)
(8,	0.477114	)
(9,	0.551244	)

    };

    \addplot[line width=0.3mm, color=ER-color,mark=triangle]  
        plot coordinates {
(0.5,	1.0000	)
(1,	0.9980	)
(2,	0.9194	)
(3,	0.8507	)
(4,	0.8149	)
(5,	0.7721	)
(6,	0.6577	)
(7,	0.5473	)
(8,	0.5164	)
(9,	0.6488	)

    };

    \addplot[line width=0.3mm, color=EP-color,mark=o]  
        plot coordinates {
(0.5,	1.0000	)
(1,	1	)
(2,	0.99801	)
(3,	0.99403	)
(4,	0.99602	)
(5,	0.99602	)
(6,	0.99403	)
(7,	0.99005	)
(8,	0.99204	)
(9,	0.975124	)

    };
    
    \addplot[line width=0.3mm, color=EK-color,mark=square]  
        plot coordinates {
(0.5,	1.0000	)
(1,	0.775124	)
(2,	0.776119	)
(3,	0.772139	)
(4,	0.808955	)
(5,	0.826866	)
(6,	0.861692	)
(7,	0.909453	)
(8,	0.943284	)
(9,	0.974129	)

    };

    \addplot[line width=0.3mm, color=GTOM-color,mark=+]  
        plot coordinates {
(0.5,	1	)
(1,	1	)
(2,	0.99801	)
(3,	0.99005	)
(4,	0.99403	)
(5,	0.99602	)
(6,	0.99005	)
(7,	0.99204	)
(8,	0.991045	)
(9,	0.979104	)
    };

    \addplot[line width=0.3mm, color=violet,mark=star]  
        plot coordinates {
(1,	0	)
(2,	0	)
(3,	0	)
(4,	0	)
(5,	0	)
(6,	0	)
(7,	0	)
(8,	0	)
(9,	0	)

    };

    \addplot[line width=0.3mm, color=BDRC-color,mark=pentagon]  
        plot coordinates {
(0.5,	1.0000	)
(1,	0.9980	)
(2,	0.8886	)
(3,	0.8378	)
(4,	0.7801	)
(5,	0.6806	)
(6,	0.5642	)
(7,	0.5811	)
(8,	0.5144	)
(9,	0.6677	)

    };

    \addplot[line width=0.3mm, color=green,mark=asterisk]  
        plot coordinates {
(1,	0	)
(2,	0	)
(3,	0	)
(4,	0	)
(5,	0	)
(6,	0	)
(7,	0	)
(8,	0	)
(9,	0	)

    };

    \addplot[line width=0.3mm, color=EOCH-color,mark=star]  
        plot coordinates {
(0.5,	1	)
(1,	1	)
(2,	0.99801	)
(3,	0.995025	)
(4,	0.991045	)
(5,	0.99204	)
(6,	0.98806	)
(7,	0.99204	)
(8,	0.99204	)
(9,	0.99005	)

    };
    \end{axis}
\end{tikzpicture}\hspace{2mm}\label{fig:node-cluster-mask-A}%
}%
\subfloat[ {\em Facebook} ]{
\begin{tikzpicture}[scale=1,every mark/.append style={mark size=2pt}]
    \begin{axis}[
        height=\columnwidth/2.4,
        width=\columnwidth/1.9,
        ylabel={\it ACC},
        xmin=0, xmax=9.5,
        ymin=0.4, ymax=1.0,
        xtick={1,3,5,7,9},
        xticklabel style = {font=\scriptsize},
        yticklabel style = {font=\small},
        xticklabels={10\%,30\%,50\%,70\%,90\%},
        every axis y label/.style={font=\small,at={(current axis.north west)},right=3mm,above=0mm},
        legend style={fill=none,font=\small,at={(0.02,0.99)},anchor=north west,draw=none},
    ]
    \addplot[line width=0.3mm, color=EB-color,mark=diamond]  
        plot coordinates {
(0.5,	1	)
(1,	0.931666	)
(2,	0.891805	)
(3,	0.875217	)
(4,	0.88413	)
(5,	0.835355	)
(6,	0.659322	)
(7,	0.692993	)
(8,	0.468433	)
(9,	0.401832	)

    };

    \addplot[line width=0.3mm, color=ER-color,mark=triangle]  
        plot coordinates {
(0.5,	1	)
(1,	0.9681	)
(2,	0.9483	)
(3,	0.9166	)
(4,	0.8911	)
(5,	0.8631	)
(6,	0.8391	)
(7,	0.8618	)
(8,	0.9611	)
(9,	0.9663	)

    };

    \addplot[line width=0.3mm, color=EP-color,mark=o]  
        plot coordinates {
(0.5,	1	)
(1,	0.984402	)
(2,	0.994306	)
(3,	0.988116	)
(4,	0.989106	)
(5,	0.990344	)
(6,	0.98465	)
(7,	0.987621	)
(8,	0.981679	)
(9,	0.978212	)

    };
    
    \addplot[line width=0.3mm, color=EK-color,mark=square]  
        plot coordinates {
(0.5,	1	)
(1,	0.979203	)
(2,	0.98465	)
(3,	0.982174	)
(4,	0.983907	)
(5,	0.983659	)
(6,	0.983907	)
(7,	0.983907	)
(8,	0.983659	)
(9,	0.980688	)

    };

    \addplot[line width=0.3mm, color=GTOM-color,mark=+]  
        plot coordinates {
(0.5,	1	)
(1,	0.996286	)
(2,	0.997029	)
(3,	0.996039	)
(4,	0.994553	)
(5,	0.995048	)
(6,	0.993563	)
(7,	0.98564	)
(8,	0.838079	)
(9,	0.787571	)

    };

    \addplot[line width=0.3mm, color=violet,mark=star]  
        plot coordinates {
(1,	0	)
(2,	0	)
(3,	0	)
(4,	0	)
(5,	0	)
(6,	0	)
(7,	0	)
(8,	0	)
(9,	0	)

    };

    \addplot[line width=0.3mm, color=BDRC-color,mark=pentagon]  
        plot coordinates {
(0.5,	1	)
(1,	0.9881	)
(2,	0.9554	)
(3,	0.8980	)
(4,	0.8789	)
(5,	0.9072	)
(6,	0.8846	)
(7,	0.8849	)
(8,	0.8765	)
(9,	0.9735	)
    };

    \addplot[line width=0.3mm, color=green,mark=asterisk]  
        plot coordinates {
(1,	0	)
(2,	0	)
(3,	0	)
(4,	0	)
(5,	0	)
(6,	0	)
(7,	0	)
(8,	0	)
(9,	0	)

    };

    \addplot[line width=0.3mm, color=EOCH-color,mark=star]  
        plot coordinates {
(0.5,	1	)
(1,	0.975241	)
(2,	0.972615	)
(3,	0.981926	)
(4,	0.995791	)
(5,	0.994306	)
(6,	0.996286	)
(7,	0.997029	)
(8,	0.996781	)
(9,	0.987126	)

    };
    \end{axis}
\end{tikzpicture}\hspace{2mm}\label{fig:node-cluster-mask-B}%
}%
\end{small}
 \vspace{-2mm}
\caption{Node clustering accuracy via spectral clustering.} \label{fig:node-cluster-mask}
\vspace{-4mm}
\end{figure}

\begin{figure}[!t]
\vspace{-2mm}
\centering
\begin{small}
\subfloat[{\em PPI}]{
\begin{tikzpicture}[scale=1,every mark/.append style={mark size=2pt}]
    \begin{axis}[
        height=\columnwidth/2.4,
        width=\columnwidth/1.9,
        ylabel={\it Micro-F1},
        xmin=0, xmax=9.5,
        ymin=0.06, ymax=0.22,
        xtick={1,3,5,7,9},
        xticklabel style = {font=\scriptsize},
        yticklabel style = {font=\small},
        xticklabels={10\%,30\%,50\%,70\%,90\%},
        every axis y label/.style={font=\small,at={(current axis.north west)},right=6mm,above=0mm},
        legend style={fill=none,font=\small,at={(0.02,0.99)},anchor=north west,draw=none},
    ]
    \addplot[line width=0.3mm, color=EB-color,mark=diamond]  
        plot coordinates {
(0.5,	0.2147330749	)
(1,	0.2048911423	)
(2,	0.1878914405	)
(3,	0.1896808828	)
(4,	0.1655234119	)
(5,	0.1515061139	)
(6,	0.1470325082	)
(7,	0.1300328064	)
(8,	0.1189979123	)
(9,	0.09275275872	)

    };

    \addplot[line width=0.3mm, color=ER-color,mark=triangle]  
        plot coordinates {
(0.5,	0.2147330749	)
(1,	0.2150313152	)
(2,	0.2094	)
(3,	0.2043	)
(4,	0.1995	)
(5,	0.1977	)
(6,	0.1885	)
(7,	0.1775	)
(8,	0.1631	)
(9,	0.1271	)

    };

    \addplot[line width=0.3mm, color=EP-color,mark=o]  
        plot coordinates {
(0.5,	0.2147330749	)
(1,	0.2138383537	)
(2,	0.2072770653	)
(3,	0.2104348345	)
(4,	0.1950492097	)
(5,	0.1932597674	)
(6,	0.1735759022	)
(7,	0.1535937966	)
(8,	0.1419624217	)
(9,	0.1127348643	)

    };
    
    \addplot[line width=0.3mm, color=EK-color,mark=square]  
        plot coordinates {
(0.5,	0.2147330749	)
(1,	0.1598568446	)
(2,	0.1395764987	)
(3,	0.1198926335	)
(4,	0.1019982106	)
(5,	0.09394572025	)
(6,	0.07903370116	)
(7,	0.07545481658	)
(8,	0.06382344169	)
(9,	0.06024455711	)

    };

    \addplot[line width=0.3mm, color=GTOM-color,mark=+]  
        plot coordinates {
(0.5,	0.2147330749	)
(1,	0.2168207575	)
(2,	0.2075753057	)
(3,	0.2028034596	)
(4,	0.2013122577	)
(5,	0.1849090367	)
(6,	0.1786459887	)
(7,	0.1756635848	)
(8,	0.1610498061	)
(9,	0.1345064122	)

    };

    \addplot[line width=0.3mm, color=violet,mark=star]  
        plot coordinates {
(1,	0	)
(2,	0	)
(3,	0	)
(4,	0	)
(5,	0	)
(6,	0	)
(7,	0	)
(8,	0	)
(9,	0	)

    };

    \addplot[line width=0.3mm, color=BDRC-color,mark=pentagon]  
        plot coordinates {
(0.5,	0.2147330749	)
(1,	0.2034	)
(2,	0.2079	)
(3,	0.2049	)
(4,	0.2022	)
(5,	0.1962	)
(6,	0.1861	)
(7,	0.1742	)
(8,	0.1455	)
(9,	0.1181	)

    };

    \addplot[line width=0.3mm, color=green,mark=asterisk]  
        plot coordinates {
(1,	0	)
(2,	0	)
(3,	0	)
(4,	0	)
(5,	0	)
(6,	0	)
(7,	0	)
(8,	0	)
(9,	0	)

    };

    \addplot[line width=0.3mm, color=EOCH-color,mark=star]  
        plot coordinates {
(0.5,	0.2147330749	)
(1,	0.2108559499	)
(2,	0.2165225171	)
(3,	0.2078735461	)
(4,	0.2016104981	)
(5,	0.2013122577	)
(6,	0.1908738443	)
(7,	0.1858037578	)
(8,	0.1741723829	)
(9,	0.1458395467	)

    };
    \end{axis}
\end{tikzpicture}\hspace{0mm}\label{fig:node-cluster-class-A}%
}%
\subfloat[ {\em BlogCatalog} ]{
\begin{tikzpicture}[scale=1,every mark/.append style={mark size=2pt}]
    \begin{axis}[
        height=\columnwidth/2.4,
        width=\columnwidth/1.9,
        ylabel={\it Micro-F1},
        xmin=0.5, xmax=9.5,
        ymin=0.17, ymax=0.39,
        xtick={1,3,5,7,9},
        xticklabel style = {font=\scriptsize},
        yticklabel style = {font=\small},
        xticklabels={10\%,30\%,50\%,70\%,90\%},
        every axis y label/.style={font=\small,at={(current axis.north west)},right=6mm,above=0mm},
        legend style={fill=none,font=\small,at={(0.02,0.99)},anchor=north west,draw=none},
    ]
    \addplot[line width=0.3mm, color=EB-color,mark=diamond]  
        plot coordinates {
(0.5,	0.3473094783	)
(1,	0.3478557771	)
(2,	0.3414367659	)
(3,	0.3341983065	)
(4,	0.3172630429	)
(5,	0.3022398252	)
(6,	0.2822999181	)
(7,	0.2491122644	)
(8,	0.2313575526	)
(9,	0.2044523354	)

    };

    \addplot[line width=0.3mm, color=ER-color,mark=triangle]  
        plot coordinates {
(0.5,		0.3473094783)
(1,	0.3563234089	)
(2,	0.3705271784	)
(3,	0.3788582355	)
(4,	0.3746244196	)
(5,	0.3804670855	)
(6,	0.3812865337	)
(7,	0.3825457525	)
(8,	0.355367386	)
(9,	0.3011472275	)

    };

    \addplot[line width=0.3mm, color=EP-color,mark=o]  
        plot coordinates {
(0.5,	0.3473094783	)
(1,	0.3494946736	)
(2,	0.3572794318	)
(3,	0.3624692707	)
(4,	0.3608303742	)
(5,	0.3501775471	)
(6,	0.3377492488	)
(7,	0.3165801694	)
(8,	0.2773832286	)
(9,	0.2301283802	)
    };
    
    \addplot[line width=0.3mm, color=EK-color,mark=square]  
        plot coordinates {
(0.5,	0.3473094783	)
(1,	0.3124829282	)
(2,	0.2596285168	)
(3,	0.2090958754	)
(4,	0.1968041519	)
(5,	0.177683693	)
(6,	0.1791860148	)
(7,	0.175361923	)
(8,	0.1727670036	)
(9,	0.1711281071	)

    };

    \addplot[line width=0.3mm, color=GTOM-color,mark=+]  
        plot coordinates {
(0.5,	0.3473094783	)
(1,	0.3341983065	)
(2,	0.3261403988	)
(3,	0.32327233	)
(4,	0.3145315488	)
(5,	0.2967768369	)
(6,	0.2842119639	)
(7,	0.2530729309	)
(8,	0.2349084949	)
(9,	0.2010379678	)

    };

    \addplot[line width=0.3mm, color=violet,mark=star]  
        plot coordinates {
(1,	0	)
(2,	0	)
(3,	0	)
(4,	0	)
(5,	0	)
(6,	0	)
(7,	0	)
(8,	0	)
(9,	0	)

    };

    \addplot[line width=0.3mm, color=BDRC-color,mark=pentagon]  
        plot coordinates {
(0.5,		0.3473094783)
(1,	0.3580988801	)
(2,	0.3721660748	)
(3,	0.3710734772	)
(4,	0.377219339	)
(5,	0.3832286261	)
(6,	0.3807702813	)
(7,	0.3724392243	)
(8,	0.3470363289	)
(9,	0.2873531822	)

    };

    \addplot[line width=0.3mm, color=green,mark=asterisk]  
        plot coordinates {
(1,	0	)
(2,	0	)
(3,	0	)
(4,	0	)
(5,	0	)
(6,	0	)
(7,	0	)
(8,	0	)
(9,	0	)

    };

    \addplot[line width=0.3mm, color=EOCH-color,mark=star]  
        plot coordinates {
(0.5,	0.3473094783	)
(1,	0.3531821907	)
(2,	0.3643813166	)
(3,	0.3761267413	)
(4,	0.3769461896	)
(5,	0.3799508331	)
(6,	0.3802239825	)
(7,	0.377219339	)
(8,	0.3684785578	)
(9,	0.3408904671	)
    };
    \end{axis}
\end{tikzpicture}\hspace{2mm}\label{fig:node-class-mask-B}%
}%
\end{small}
 \vspace{-2mm}
\caption{Node classification accuracy via node2vec.} \label{fig:node-class-mask}
\vspace{-4mm}
\end{figure}
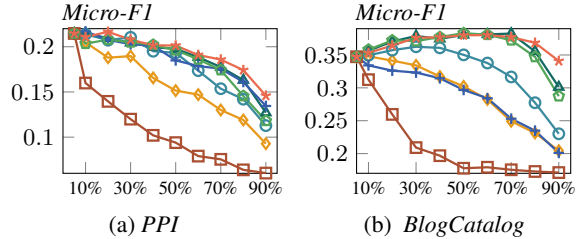

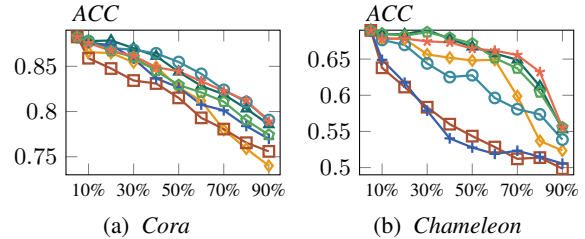
\begin{figure}[!t]
\vspace{-2mm}
\centering
\begin{small}
\subfloat[ {\em Cora} ]{
\begin{tikzpicture}[scale=1,every mark/.append style={mark size=2pt}]
    \begin{axis}[
        height=\columnwidth/2.4,
        width=\columnwidth/1.9,
        ylabel={\it ACC},
        xmin=0, xmax=9.5,
        ymin=0.73, ymax=0.89,
        xtick={1,3,5,7,9},
        xticklabel style = {font=\scriptsize},
        yticklabel style = {font=\small},
        xticklabels={10\%,30\%,50\%,70\%,90\%},
        every axis y label/.style={font=\small,at={(current axis.north west)},right=4mm,above=0mm},
        legend style={fill=none,font=\small,at={(0.02,0.99)},anchor=north west,draw=none},
    ]

    \addplot[line width=0.3mm, color=EB-color,mark=diamond]  
        plot coordinates {
(0.5,	0.8825	)
(1,	0.8655	)
(2,	0.8648	)
(3,	0.8544	)
(4,	0.8494	)
(5,	0.8266	)
(6,	0.8118	)
(7,	0.7799	)
(8,	0.7589	)
(9,	0.7399	)

    };

    \addplot[line width=0.3mm, color=ER-color,mark=triangle]  
        plot coordinates {
(0.5,	0.8825	)
(1,	0.8779	)
(2,	0.8786	)
(3,	0.8701	)
(4,	0.8613	)
(5,	0.8439	)
(6,	0.8284	)
(7,	0.8190	)
(8,	0.8031	)
(9,	0.7856	)

    };

    \addplot[line width=0.3mm, color=EP-color,mark=o]  
        plot coordinates {
(0.5,	0.8825	)
(1,	0.878	)
(2,	0.8703	)
(3,	0.8683	)
(4,	0.8644	)
(5,	0.8548	)
(6,	0.8417	)
(7,	0.824	)
(8,	0.8105	)
(9,	0.7906	)

    };
    
    \addplot[line width=0.3mm, color=EK-color,mark=square]  
        plot coordinates {
(0.5,	0.8825	)
(1,	0.859	)
(2,	0.8476	)
(3,	0.8343	)
(4,	0.831	)
(5,	0.8151	)
(6,	0.793	)
(7,	0.7806	)
(8,	0.7655	)
(9,	0.7559	)

    };

    \addplot[line width=0.3mm, color=GTOM-color,mark=+]  
        plot coordinates {
(0.5,	0.8825	)
(1,	0.8758	)
(2,	0.8736	)
(3,	0.8585	)
(4,	0.8367	)
(5,	0.8266	)
(6,	0.8076	)
(7,	0.8013	)
(8,	0.7839	)
(9,	0.7697	)

    };

    \addplot[line width=0.3mm, color=violet,mark=star]  
        plot coordinates {
(1,	0	)
(2,	0	)
(3,	0	)
(4,	0	)
(5,	0	)
(6,	0	)
(7,	0	)
(8,	0	)
(9,	0	)

    };

    \addplot[line width=0.3mm, color=BDRC-color,mark=pentagon]  
        plot coordinates {
(0.5,	0.8825	)
(1,	0.8777	)
(2,	0.8670	)
(3,	0.8603	)
(4,	0.8450	)
(5,	0.8293	)
(6,	0.8212	)
(7,	0.8109	)
(8,	0.7904	)
(9,	0.7742	)

    };

    \addplot[line width=0.3mm, color=green,mark=asterisk]  
        plot coordinates {
(1,	0	)
(2,	0	)
(3,	0	)
(4,	0	)
(5,	0	)
(6,	0	)
(7,	0	)
(8,	0	)
(9,	0	)

    };

    \addplot[line width=0.3mm, color=EOCH-color,mark=star]  
        plot coordinates {
(0.5,	0.8825	)
(1,	0.8756	)
(2,	0.8673	)
(3,	0.8616	)
(4,	0.8491	)
(5,	0.8445	)
(6,	0.8336	)
(7,	0.8227	)
(8,	0.8124	)
(9,	0.7889	)

    };
    \end{axis}
\end{tikzpicture}\hspace{0mm}\label{fig:node-class-mask-C}%
}%
\subfloat[ {\em Chameleon} ]{
\begin{tikzpicture}[scale=1,every mark/.append style={mark size=2pt}]
    \begin{axis}[
        height=\columnwidth/2.4,
        width=\columnwidth/1.9,
        ylabel={\it ACC},
        xmin=0, xmax=9.5,
        ymin=0.49, ymax=0.69,
        xtick={1,3,5,7,9},
        xticklabel style = {font=\scriptsize},
        yticklabel style = {font=\small},
        xticklabels={10\%,30\%,50\%,70\%,90\%},
        every axis y label/.style={font=\small,at={(current axis.north west)},right=4mm,above=0mm},
        legend style={fill=none,font=\small,at={(0.02,0.99)},anchor=north west,draw=none},
    ]
    \addplot[line width=0.3mm, color=EB-color,mark=diamond]  
        plot coordinates {
(0.5,	0.6897	)
(1,	0.6789	)
(2,	0.6778	)
(3,	0.6565	)
(4,	0.6519	)
(5,	0.6481	)
(6,	0.649	)
(7,	0.5987	)
(8,	0.5369	)
(9,	0.5235	)

    };

    \addplot[line width=0.3mm, color=ER-color,mark=triangle]  
        plot coordinates {
(0.5,	0.6897	)
(1,	0.6844	)
(2,	0.6844	)
(3,	0.6905	)
(4,	0.6778	)
(5,	0.6670	)
(6,	0.6578	)
(7,	0.6490	)
(8,	0.6125	)
(9,	0.5525	)

    };

    \addplot[line width=0.3mm, color=EP-color,mark=o]  
        plot coordinates {
(0.5,	0.6897	)
(1,	0.6765	)
(2,	0.6697	)
(3,	0.644	)
(4,	0.6251	)
(5,	0.6277	)
(6,	0.5963	)
(7,	0.5807	)
(8,	0.5736	)
(9,	0.5387	)

    };
    
    \addplot[line width=0.3mm, color=EK-color,mark=square]  
        plot coordinates {
(0.5,	0.6897	)
(1,	0.638	)
(2,	0.6114	)
(3,	0.5835	)
(4,	0.56	)
(5,	0.5435	)
(6,	0.5284	)
(7,	0.5121	)
(8,	0.5138	)
(9,	0.4985	)

    };

    \addplot[line width=0.3mm, color=GTOM-color,mark=+]  
        plot coordinates {
(0.5,	0.6897	)
(1,	0.649	)
(2,	0.6174	)
(3,	0.5787	)
(4,	0.5404	)
(5,	0.5277	)
(6,	0.5189	)
(7,	0.5233	)
(8,	0.5149	)
(9,	0.5055	)

    };

    \addplot[line width=0.3mm, color=violet,mark=star]  
        plot coordinates {
(1,	0	)
(2,	0	)
(3,	0	)
(4,	0	)
(5,	0	)
(6,	0	)
(7,	0	)
(8,	0	)
(9,	0	)

    };

    \addplot[line width=0.3mm, color=BDRC-color,mark=pentagon]  
        plot coordinates {
(0.5,	0.6897	)
(1,	0.6853	)
(2,	0.6824	)
(3,	0.6873	)
(4,	0.6796	)
(5,	0.6716	)
(6,	0.6516	)
(7,	0.6371	)
(8,	0.6040	)
(9,	0.5556	)

    };

    \addplot[line width=0.3mm, color=green,mark=asterisk]  
        plot coordinates {
(1,	0	)
(2,	0	)
(3,	0	)
(4,	0	)
(5,	0	)
(6,	0	)
(7,	0	)
(8,	0	)
(9,	0	)

    };

    \addplot[line width=0.3mm, color=EOCH-color,mark=star]  
        plot coordinates {
(0.5,	0.6897	)
(1,	0.678	)
(2,	0.6785	)
(3,	0.6745	)
(4,	0.6732	)
(5,	0.6644	)
(6,	0.6618	)
(7,	0.6556	)
(8,	0.6321	)
(9,	0.5552	)

    };
    \end{axis}
\end{tikzpicture}\hspace{2mm}\label{fig:node-class-mask-D}%
}%
\end{small}
 \vspace{-3mm}
\caption{Node classification accuracy via GCN.} \label{fig:node-semi-class-mask}
\vspace{-6mm}
\end{figure}


\pgfplotsset{ every non boxed x axis/.append style={x axis line style=-} }
\pgfplotsset{ every non boxed y axis/.append style={y axis line style=-} }
\begin{figure*}[!t]
\centering
\begin{small}
\begin{tikzpicture}
    \begin{customlegend}[
        legend entries={EB, ER, EP, EK, GTOM, BDRC, \algo},
        legend columns=7,
        area legend,
        legend style={at={(0.45,1.15)},anchor=north,draw=none,font=\small,column sep=0.15cm}]
        \addlegendimage{preaction={fill, EB-color}, pattern=dots}    
        \addlegendimage{preaction={fill, ER-color}, pattern={north east lines}} 
        \addlegendimage{preaction={fill, EP-color}, pattern=crosshatch dots} 
        \addlegendimage{preaction={fill, EK-color}, pattern={grid}}    
        \addlegendimage{preaction={fill, GTOM-color}, pattern=crosshatch}    
        \addlegendimage{preaction={fill, BDRC-color}, pattern=north west lines}    
        \addlegendimage{preaction={fill, EOCH-color}}    
    \end{customlegend}
\end{tikzpicture}
\\[-\lineskip]
\vspace{-5mm}
\subfloat[{\em Cora}]{
\begin{tikzpicture}[scale=1]
\begin{axis}[
        height=\columnwidth/2.6,
        width=\columnwidth/2.25,
    xtick=\empty,
    ybar=1.5pt,
    bar width=0.2cm,
    enlarge x limits=true,
    ylabel={\em time} (sec),
    xticklabel=\empty,
    yticklabel style = {font=\scriptsize},
    ymin=0.1,
    ymax=200,
    ytick={0.1,1,10,100},
    yticklabels={$0.1$,$1$,$10$,$10^2$},
    ymode=log,
    log origin y=infty,
    log basis y={10},
    every axis y label/.style={at={(current axis.north west)},right=6mm,above=0mm},
    legend style={at={(0.02,0.98)},anchor=north west,cells={anchor=west},font=\tiny}
    ]

\addplot [preaction={fill, EB-color}, pattern=dots] coordinates {(1,114.225753) }; %
\addplot [preaction={fill, ER-color}, pattern={north east lines}] coordinates {(1,2.888346) }; %
\addplot [preaction={fill, EP-color}, pattern=crosshatch dots] coordinates {(1,0.283371) }; %
\addplot [preaction={fill, EK-color}, pattern={grid}] coordinates {(1,0.281046) }; %
\addplot [preaction={fill, GTOM-color}, pattern=crosshatch] coordinates {(1,4.159003) }; %
\addplot [preaction={fill, BDRC-color}, pattern=north west lines] coordinates {(1,2.572407) }; %
\addplot [preaction={fill, EOCH-color}] coordinates {(1,0.16211938858032227) }; %
\end{axis}
\end{tikzpicture}\hspace{0mm}\label{fig:time-cora}%
}%
\subfloat[{\em Email-EU}]{
\begin{tikzpicture}[scale=1]
\begin{axis}[
        height=\columnwidth/2.6,
        width=\columnwidth/2.25,
    xtick=\empty,
    ybar=1.5pt,
    bar width=0.2cm,
    enlarge x limits=true,
    ylabel={\em time} (sec),
    xticklabel=\empty,
    yticklabel style = {font=\scriptsize},
    ymin=0.1,
    ymax=100,
    ytick={0.1,1,10,100},
    yticklabels={$0.1$,$1$,$10$,$10^2$},
    log origin y=infty,
    ymode=log,
    log basis y={10},
    every axis y label/.style={at={(current axis.north west)},right=6mm,above=0mm},
    legend style={at={(0.02,0.98)},anchor=north west,cells={anchor=west},font=\tiny}
    ]

\addplot [preaction={fill, EB-color}, pattern=dots] coordinates {(1,87.489527) }; %
\addplot [preaction={fill, ER-color}, pattern={north east lines}] coordinates {(1,0.652397) }; %
\addplot [preaction={fill, EP-color}, pattern=crosshatch dots] coordinates {(1,0.926601) }; %
\addplot [preaction={fill, EK-color}, pattern={grid}] coordinates {(1,0.932571) }; %
\addplot [preaction={fill, GTOM-color}, pattern=crosshatch] coordinates {(1,15.763865) }; %
\addplot [preaction={fill, BDRC-color}, pattern=north west lines] coordinates {(1,0.617957) }; %
\addplot [preaction={fill, EOCH-color}] coordinates {(1,0.45749545097351074) }; %
\end{axis}
\end{tikzpicture}\hspace{0mm}\label{fig:time-citeseer}%
}%
\subfloat[{\em Chameleon}]{
\begin{tikzpicture}[scale=1]
\begin{axis}[
        height=\columnwidth/2.6,
        width=\columnwidth/2.25,
    xtick=\empty,
    ybar=1.5pt,
    bar width=0.2cm,
    enlarge x limits=true,
    ylabel={\em time} (sec),
    xticklabel=\empty,
    yticklabel style = {font=\scriptsize},
    ymin=0.1,
    ymax=300,
    ytick={0.1,1,10,100},
    yticklabels={$0.1$,$1$,$10$,$10^2$},
    ymode=log,
    log origin y=infty,
    log basis y={10},
    every axis y label/.style={at={(current axis.north west)},right=6mm,above=0mm},
    legend style={at={(0.02,0.98)},anchor=north west,cells={anchor=west},font=\tiny}
    ]

\addplot [preaction={fill, EB-color}, pattern=dots] coordinates {(1,261.322013) }; %
\addplot [preaction={fill, ER-color}, pattern={north east lines}] coordinates {(1,2.356988) }; %
\addplot [preaction={fill, EP-color}, pattern=crosshatch dots] coordinates {(1,1.349415) }; %
\addplot [preaction={fill, EK-color}, pattern={grid}] coordinates {(1,1.347593) }; %
\addplot [preaction={fill, GTOM-color}, pattern=crosshatch] coordinates {(1,34.506548) }; %
\addplot [preaction={fill, BDRC-color}, pattern=north west lines] coordinates {(1,1.820206) }; %
\addplot [preaction={fill, EOCH-color}] coordinates {(1,0.5426535606384277) }; %
\end{axis}
\end{tikzpicture}\hspace{0mm}\label{fig:time-blogcatalog}%
}%
\subfloat[{\em PPI}]{
\begin{tikzpicture}[scale=1]
\begin{axis}[
        height=\columnwidth/2.6,
        width=\columnwidth/2.25,
    xtick=\empty,
    yticklabel style = {font=\scriptsize},
    ybar=1.5pt,
    bar width=0.2cm,
    enlarge x limits=true,
    ylabel={\em time} (sec),
    xticklabel=\empty,
    ymin=0.1,
    ymax=1000,
    ytick={0.1,1,10,100,1000},
    yticklabels={$0.1$,$1$,$10$,$10^2$,$10^3$},
    ymode=log,
    log origin y=infty,
    log basis y={10},
    every axis y label/.style={at={(current axis.north west)},right=6mm,above=0mm},
    legend style={at={(0.02,0.98)},anchor=north west,cells={anchor=west},font=\tiny}
    ]

\addplot [preaction={fill, EB-color}, pattern=dots] coordinates {(1,867.020077) }; %
\addplot [preaction={fill, ER-color}, pattern={north east lines}] coordinates {(1,7.310528) }; %
\addplot [preaction={fill, EP-color}, pattern=crosshatch dots] coordinates {(1,1.6801016330718994) }; %
\addplot [preaction={fill, EK-color}, pattern={grid}] coordinates {(1,1.6652047634124756) }; %
\addplot [preaction={fill, GTOM-color}, pattern=crosshatch] coordinates {(1,33.391757) }; %
\addplot [preaction={fill, BDRC-color}, pattern=north west lines] coordinates {(1,6.896416) }; %
\addplot [preaction={fill, EOCH-color}] coordinates {(1,0.47753334045410156) }; %
\end{axis}
\end{tikzpicture}\hspace{0mm}\label{fig:time-flickr}%
}%
\subfloat[{\em Facebook}]{
\begin{tikzpicture}[scale=1]
\begin{axis}[
        height=\columnwidth/2.6,
        width=\columnwidth/2.25,
    xtick=\empty,
    ybar=1.5pt,
    bar width=0.2cm,
    enlarge x limits=true,
    ylabel={\em time} (sec),
    xticklabel=\empty,
    yticklabel style = {font=\scriptsize},
    ymin=1.0,
    ymax=2000,
    ytick={1,10,100,1000},
    yticklabels={$1$,$10$,$10^2$,$10^3$},
    ymode=log,
    log origin y=infty,
    log basis y={10},
    every axis y label/.style={at={(current axis.north west)},right=6mm,above=0mm},
    legend style={at={(0.02,0.98)},anchor=north west,cells={anchor=west},font=\tiny}
    ]

\addplot [preaction={fill, EB-color}, pattern=dots] coordinates {(1,1548.415765) }; %
\addplot [preaction={fill, ER-color}, pattern={north east lines}] coordinates {(1,8.761945) }; %
\addplot [preaction={fill, EP-color}, pattern=crosshatch dots] coordinates {(1,3.672628) }; %
\addplot [preaction={fill, EK-color}, pattern={grid}] coordinates {(1,4.219920) }; %
\addplot [preaction={fill, GTOM-color}, pattern=crosshatch] coordinates {(1,80.095309) }; %
\addplot [preaction={fill, BDRC-color}, pattern=north west lines] coordinates {(1,8.799332) }; %
\addplot [preaction={fill, EOCH-color}] coordinates {(1,1.2584855556488037) }; %
\end{axis}
\end{tikzpicture}\hspace{0mm}\label{fig:time-pubmed}%
}%
\subfloat[{\em BlogCatalog}]{
\begin{tikzpicture}[scale=1]
\begin{axis}[
        height=\columnwidth/2.6,
        width=\columnwidth/2.25,
    xtick=\empty,
    ybar=1.5pt,
    bar width=0.2cm,
    enlarge x limits=true,
    ylabel={\em time} (sec),
    xticklabel=\empty,
    yticklabel style = {font=\scriptsize},
    ymin=1,
    ymax=20000,
    ymode=log,
    ytick={1,10,100,1000,10000},
    yticklabels={$1$,$10$,$10^2$,$10^3$,$10^4$},
    log basis y={10},
    every axis y label/.style={at={(current axis.north west)},right=6mm,above=0mm},
    legend style={at={(0.02,0.98)},anchor=north west,cells={anchor=west},font=\tiny}
    ]

\addplot [preaction={fill, EB-color}, pattern=dots] coordinates {(1,18465.36659) }; %
\addplot [preaction={fill, ER-color}, pattern={north east lines}] coordinates {(1,1031.124188) }; %
\addplot [preaction={fill, EP-color}, pattern=crosshatch dots] coordinates {(1,9.499058485031128) }; %
\addplot [preaction={fill, EK-color}, pattern={grid}] coordinates {(1,9.280590057373047) }; %
\addplot [preaction={fill, GTOM-color}, pattern=crosshatch] coordinates {(1,312.439891) }; %
\addplot [preaction={fill, BDRC-color}, pattern=north west lines] coordinates {(1,116.850831) }; %
\addplot [preaction={fill, EOCH-color}] coordinates {(1,4.464556694030762) }; %
\end{axis}
\end{tikzpicture}\hspace{2mm}\label{fig:time-corafull}%
}%
\vspace{-3mm}
\end{small}
\caption{Computation time in seconds.} \label{fig:time}
\vspace{-2mm}
\end{figure*}
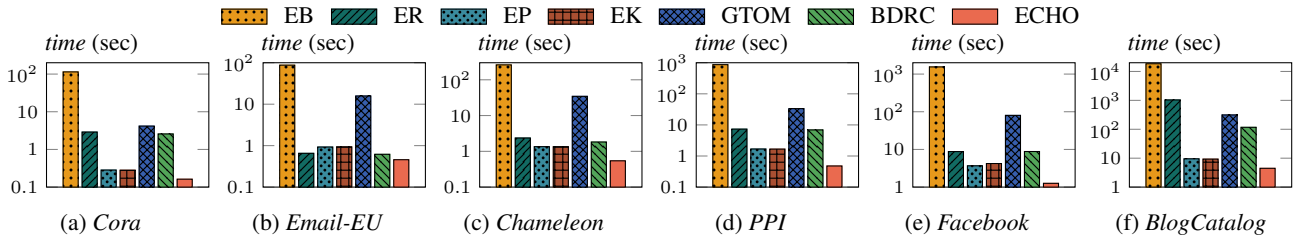

\begin{figure}[!t]
\centering
\begin{small}
\begin{tikzpicture}
    \begin{customlegend}[legend columns=3,
        legend entries={Simple ISM, Adaptive ISM, DEV Method},
        legend style={at={(0.45,1.35)},anchor=north,draw=none,font=\small,column sep=0.2cm}]
    \addlegendimage{line width=0.2mm,mark size=4pt,color=EOCH-color,mark=star}
    \addlegendimage{line width=0.2mm,mark size=4pt,color=EB-color,mark=o}
    \addlegendimage{line width=0.2mm,mark size=4pt,color=cyan,mark=square}
    \end{customlegend}
\end{tikzpicture}
\\[-\lineskip]
\vspace{-4mm}
\subfloat[{\em PPI}]{
\begin{tikzpicture}[scale=1,every mark/.append style={mark size=2pt}]
    \begin{axis}[
        height=\columnwidth/2.5,
        width=\columnwidth/1.9,
        ylabel={\it Average Error},
        xmin=0.5, xmax=7.5,
        ymin=0, ymax=0.1,
        xtick={1,3,5,7},
        xticklabel style = {font=\tiny},
        yticklabel style = {font=\tiny},
        xticklabels={1e-1,1e-5,1e-20,1e-40},
        ymode=log,
        log basis y={10},
        every axis y label/.style={font=\small,at={(current axis.north west)},right=8mm,above=0mm},
        legend style={fill=none,font=\small,at={(0.02,0.99)},anchor=north west,draw=none},
    ]

        \addplot[line width=0.3mm, color=EOCH-color,mark=star]  
        plot coordinates {
(1,	0.006328825687	)
(2,	0.0007911032109	)
(3,	7.73E-07	)
(4,	5.89E-12	)
(5,	5.04E-18	)
(6,	5.04E-18	)
(7,	0	)

    };

    \addplot[line width=0.3mm, color=EB-color,mark=o]  
        plot coordinates {
(1,	0.006328825687	)
(2,	0.0007911032109	)
(3,	7.73E-07	)
(4,	5.89E-12	)
(5,	2.60E-17	)
(6,	2.60E-17	)
(7,	2.60E-17	)

    };

    \addplot[line width=0.3mm, color=cyan,mark=square]  
        plot coordinates {
(1,	0.0002791210942	)
(2,	1.46E-05	)
(3,	1.81E-09	)
(4,	7.82E-16	)
(5,	9.98E-18	)
(6,	5.65E-18	)
(7,	9.98E-19	)

    };
    \end{axis}
\end{tikzpicture}\hspace{0mm}\label{fig:eps-ppi}%
}%
\subfloat[ {\em BlogCatalog} ]{
\begin{tikzpicture}[scale=1,every mark/.append style={mark size=2pt}]
    \begin{axis}[
        height=\columnwidth/2.5,
        width=\columnwidth/1.9,
        ylabel={\it Average Error},
        xmin=0.5, xmax=7.5,
        ymin=0, ymax=0.1,
        xtick={1,3,5,7},
        xticklabel style = {font=\tiny},
        yticklabel style = {font=\tiny},
        xticklabels={1e-1,1e-5,1e-20,1e-40},
        ymode=log,
        log basis y={10},
        every axis y label/.style={font=\small,at={(current axis.north west)},right=8mm,above=0mm},
        legend style={fill=none,font=\small,at={(0.02,0.99)},anchor=north west,draw=none},
    ]

    \addplot[line width=0.3mm, color=EOCH-color,mark=star]  
        plot coordinates {
(1,	0.002668418298	)
(2,	0.0003335522872	)
(3,	3.26E-07	)
(4,	2.49E-12	)
(5,	6.65E-17	)
(6,	5.20E-18	)
(7,	5.20E-18	)

    };

    \addplot[line width=0.3mm, color=EB-color,mark=o]  
        plot coordinates {
(1,	0.005336836595	)
(2,	0.001334209149	)
(3,	2.61E-06	)
(4,	3.98E-11	)
(5,	1.35E-17	)
(6,	1.35E-17	)
(7,	1.35E-17	)

    };

    \addplot[line width=0.3mm, color=cyan,mark=square]  
        plot coordinates {
(1,	7.29E-05	)
(2,	2.35E-06	)
(3,	1.26E-10	)
(4,	1.66E-17	)
(5,	4.31E-18	)
(6,	5.35E-18	)
(7,	4.31E-19	)

    };
    \end{axis}
\end{tikzpicture}\hspace{0mm}\label{fig:eps-blog}%
}%
\end{small}
 \vspace{-3mm}
\caption{The convergence when varying $\epsilon$.} \label{fig:epsilon}
\vspace{-4mm}
\end{figure}
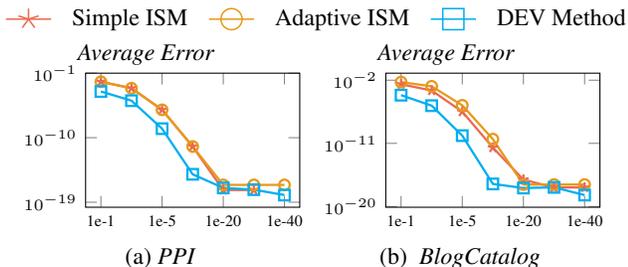

\stitle{Node Clustering} First, we apply spectral clustering on the full datasets to generate the clustering results as ground-truth node labels. Fig. \ref{fig:node-cluster-mask} depicts the accuracy scores achieved by all the 7 edge centrality measures on {\em Email-EU} and {\em Facebook} when varying $\rho$ from $0.1$ to $0.9$ (i.e., removing $10\%$~$90\%$ edges from the datasets). We can observe that both EP and \algo consistently attain the highest clustering quality on the two datasets, whereas GTOM and EK are among the best on one dataset but suffer from conspicuous performance degradation on the other one. 


\stitle{Node Classification with node2vec} Fig. \ref{fig:node-class-mask} reports the micro-F1 scores for multi-label node classification by feeding the node embeddings obtained on the residual networks $G^{\prime}$ to the one-vs-rest logistic regression classifier when varying $\rho$ from $0.1$ to $0.9$.  From Fig. \ref{fig:node-class-mask}, we can make the following observations. First, \algo consistently obtained the best performance on {\em PPI} and {\em BlogCatalog}. Particularly, when $\rho=0.9$ (i.e., $90\%$ edges are removed), \algo is superior to all the competitors by at least $1.13\%$ and $14\%$ gain in micro-F1 on {\em PPI} and {\em BlogCatalog}, respectively. Second, it can be observed that EP exhibits radically different behaviors in this task, which is even inferior to BDRC and ER.

\stitle{Semi-supervised Node Classification with GCN} Fig. \ref{fig:node-semi-class-mask}, we present the semi-supervised node classification accuracy scores with GCN over the residuals networks created by varying $\rho$ from $0.1$ to $0.9$. on {\em Cora}, we can see that \algo is second to ER and EP when $\rho\le 70\%$. By contrast, when $\rho \ge 70\%$, \algo is on par with the best competitor EP. Similar observation can be made on {\em Chameleon}, where \algo is slightly inferior to ER and BDRC when $10\%\le \rho \le 50\%$ but obtains the best performance when $\rho\ge 60\%$. It's worth noting that on {\em Chameleon}, EP produces a poor performance. This phenomenon reveals that EP is sensitive to both datasets and graph analytics tasks.  

In summary, existing edge centrality measures either fall short of capturing network topology accurately or are sensitive to network types and structures. In comparison, \algo achieves superior and stable performance in terms of diverse tasks on various networks, corrobarating its effectiveness and robustness in the accurate identification of important and unimportant edges.

\subsection{Efficiency Evaluation}

In Fig. \ref{fig:time}, we display the time costs required for the computation of \algo and the other six centrality measures of all edges in the six tested datasets. Note that the $y$-axis is in log-scale, and the measurement unit for running time is a second (sec). We can observe that \algo consistently consumes the least computation time on all datasets, whereas EB is the most expensive due to its quadratic time complexity. 
Compared to the most efficient existing metrics EP and EK, \algo is $2\times- 3.49\times$ faster on most datasets. As for ER and BDRC who are strong competitors in Fig. \ref{fig:node-class-mask} and Fig. \ref{fig:node-semi-class-mask}, \algo can be up to orders of magnitude faster.
\eat{
The efficiency of ER and BDRC is comparable to that of EP, EK, and \algo on {\em Email-EU} and {\em Chameleon}. 
The reason is that the EP, EK, and \algo involve hundreds of iterations to converge, while the computations of ER and BDRC mainly rely on the matrix inverse, which can be done efficiently in networks with a small number of nodes or edges. 
}


\subsection{Convergence Study}
In this set of experiments, we empirically study the convergence of the three algorithms in Section \ref{sec:algos} for \algo estimation. Let $\zm$ by Algo. \ref{alg:algo} with $t=150$ be the exact \algo vector and $\zm^{\prime}$ be the estimated \algo vector. We define the {\em average error} of the estimation $\zm^{\prime}$ as $\frac{1}{m}\|\zm-\zm^{\prime}\|_1$. 
Fig. \ref{fig:epsilon} illustrates the average errors by the Algo. \ref{alg:algo}, Algo. \ref{alg:algo2}, and Algo. \ref{alg:algo3} on representative datasets {\em PPI} and {\em BlogCatalog} when varying $\epsilon$ from $0.1$ to $10^{-40}$. For a fair comparison, the maximum number $t$ of iterations in Algo. \ref{alg:algo3} is computed as Line 4 in Algo. \ref{alg:algo}. From Fig. \ref{fig:epsilon}, we can see that the Simple ISM and Adaptive ISM have almost identical convergence behaviors, whereas the DEV method presents conspicuously faster convergence than the iterative summation methods.

\section{Other Related Work}



Centrality is a major focus of network analysis, for which there exists a large body of literature, as surveyed in \cite{borgatti2006graph,landherr2010critical,boldi2014axioms}. These centrality indices are mainly designed for nodes (a.k.a. vertices). 
Based on the walk structures used for characterization, most of them can be classified into radial and medial centralities \cite{borgatti2006graph}. The former count walks (or trails, paths, geodesics) that start from or end at the given node, e.g., degree centrality, closeness centrality \cite{bavelas1950communication}, eigenvalue centrality \cite{bonacich1972factoring}, and its variants Katz centrality \cite{katz1953new} and PageRank \cite{page1998pagerank}. By contrast, the latter count walks that pass through the given node, e.g., betweenness centrality \cite{freeman1977set} and flow betweenness \cite{freeman1991centrality}.
A similar categorization scheme for these centrality indices based on network flow is proposed in \cite{borgatti2005centrality}. 
\cite{saxena2020centrality} gave a detailed discussion regarding the extensions, approximation/update algorithms, and applications of these prominent measures, as well as a brief introduction to new centrality measures proposed in recent years.
\cite{bloch2023centrality} developed a new taxonomy of these metrics based on nodal statistics and offered a list of axioms to characterize them.


\section{Conclusion}
In this paper, we propose a new centrality metric \algo for ranking edges in networks. \algo is built on our carefully crafted neighborhood-based optimization objective function, which seeks to capture the local out-degree information and ensure adjacent edges consistent in edge ranking. 
We conduct thorough theoretical analyses to interpret \algo so as to facilitate an intuitive and deeper understanding of \algo. Additionally, three linear-time algorithms with rigorous theoretical assurance are devised for the approximation of \algo.
\algo overcomes the limitations of existing edge centrality measures in terms of both ranking effectiveness and computation efficiency. which is validated by our experiments on real-world datasets in various graph analytics tasks. 



\bibliography{main}

\appendix
\section{Proofs}
\begin{proof}[\bf Proof of Theorem \ref{lem:C-def}]
We rewrite the second term in Eq. \eqref{eq:obj} as 
\begin{align*}
& \sum_{e_i,e_j\in \EDG,i<j}\frac{1}{2}{\sum_{v\in e_i\cap e_j}\frac{\left(\zm_i-\zm_j\right)^2}{|\N^{+}(v)|}} \\
& =  \frac{1}{2}\left(\sum_{e_i\in \EDG}{\zm_i^2} + \sum_{e_j\in \EDG}{\zm_j^2} - 2\sum_{e_i,e_j\in \EDG}{\sum_{v\in e_i\cap e_j}\frac{\zm_i\zm_j}{2|\N^{+}(v)|}}\right) \\
& = \zm^{\top}\IM\zm - \frac{1}{2}\zm^{\top}{\EM}^{\top}{\DM^{-1}}\EM\zm = \zm^{\top}\left(\IM- \frac{1}{2}{\EM}^{\top}\DM^{-1}\EM\right)\zm.
\end{align*}
By setting the derivative of the objective function $\mathcal{L}$ to 0, i.e., $\frac{\partial \mathcal{L}}{\partial \zm} = (1-\alpha)\cdot(\zm-\xm) + \alpha\cdot \left(\IM- \frac{1}{2}{\EM}^{\top}\DM^{-1}\EM\right)\zm = 0$, we obtain
\begin{align}
& \zm = \left(\IM+\frac{\alpha}{1-\alpha}\cdot \left(\IM- \frac{1}{2}{\EM}^{\top}\DM^{-1}\EM\right)\right)^{-1}\xm \notag \\
&= (1-\alpha)\cdot\left((1-\alpha)\cdot\IM+{\alpha}\cdot \left(\IM- \frac{1}{2}{\EM}^{\top}\DM^{-1}\EM\right)\right)^{-1} \xm \notag \\
& = (1-\alpha)\cdot\left(\IM - {\alpha}\cdot \frac{1}{2}{\EM}^{\top}\DM^{-1}\EM\right)^{-1}\xm. \label{eq:z-nms}
\end{align}
Theorem \ref{lem:C-def} is then proved.
\end{proof}

\begin{proof}[\bf Proof of Lemma \ref{lem:C-def-eig}]
Let the singular value decomposition of $\frac{1}{\sqrt{2}}\EM\DM^{-\frac{1}{2}}$ be $\UM\boldsymbol{\Sigma}\VM^{\top}$. On the other hand, by the semi-unitary property of singular vectors, i.e., $\UM^{\top}\UM=\VM^{\top}\VM=\IM$, we have
\begin{align*}
& \left(\IM-\frac{{\alpha}}{2}{\EM}^{\top}\DM^{-1}\EM\right)^{-1} = \left(\IM-{\alpha}\cdot \UM\boldsymbol{\Sigma}\VM^{\top}\VM\boldsymbol{\Sigma}\UM^{\top}\right)^{-1}\\
& = \left(\IM-{\alpha}\cdot \UM\boldsymbol{\Sigma}^2\UM^{\top}\right)^{-1}  = \left(\UM\cdot(\IM-{\alpha}\boldsymbol{\Sigma}^2)\cdot\UM^{\top}\right)^{-1}\\
& = \UM\frac{1}{\IM-{\alpha}\boldsymbol{\Sigma}^2}\UM^{\top}.
\end{align*}
Plugging this equality into Eq. \eqref{eq:zm_inverse} finishes the proof.
\end{proof}

\begin{proof}[\bf Proof of Lemma \ref{lem:U-Lambda-Sigma}]
Recall that $\widetilde{\UM}$ and $\widetilde{\boldsymbol{\Lambda}}$ are the eigenvectors and eigenvalues of $\widetilde{\LM}=\IM-\frac{1}{2}{\EM}^{\top}\DM^{-1}\EM$, i.e., 
$$\IM-\frac{1}{2}{\EM}^{\top}\DM^{-1}\EM=\widetilde{\UM}\widetilde{\boldsymbol{\Lambda}}\widetilde{\UM}^{\top}.$$
Then, we get $\frac{1}{2}{\EM}^{\top}\DM^{-1}\EM=\widetilde{\UM}\cdot(\IM-\widetilde{\boldsymbol{\Lambda}})\cdot \widetilde{\UM}^{\top}$, implying that $\widetilde{\UM}$ and $\IM-\widetilde{\boldsymbol{\Lambda}}$ correspond to the eigenvectors and eigenvalues of $\frac{1}{2}{\EM}^{\top}\DM^{-1}\EM$, respectively. Recall the singular value decomposition of $\frac{1}{\sqrt{2}}{\EM}^{\top}\DM^{-\frac{1}{2}}$ is $\UM\boldsymbol{\Sigma}\VM^{\top}$. According to \cite{strang2023introduction}, $\UM$ and $\boldsymbol{\Sigma}^2$ are the eigenvectors and eigenvalues of $\frac{1}{\sqrt{2}}{\EM}^{\top}\DM^{-\frac{1}{2}}\cdot (\frac{1}{\sqrt{2}}{\EM}^{\top}\DM^{-\frac{1}{2}})^{\top}=\frac{1}{2}{\EM}^{\top}\DM^{-1}\EM$, respectively. Accordingly, we obtain that $\widetilde{\UM}=\UM$ and $\widetilde{\boldsymbol{\Lambda}}=\IM-\boldsymbol{\Sigma}^2$ and the lemma is proved.
\end{proof}

\begin{proof}[\bf Proof of Lemma \ref{lem:C-def-sum}]
As per Lemma 4.1 in \cite{denny2020bounds}, each eigenvalue $\lambda$ of a matrix ${\alpha}\cdot \frac{1}{2}{\EM}^{\top}\DM^{-1}\EM$ is bounded by
\begin{align*}
& |\lambda| \le {\alpha}\cdot \max_{e_i\in \EDG}\sum_{e_j\in \EDG}\left(\frac{1}{2}{\EM}^{\top}\DM^{-1}\EM\right)[e_i,e_j] \\
& =\alpha\cdot \max_{e_i\in \EDG}\sum_{e_j\in \EDG}{\sum_{v\in e_i\cap e_j}\frac{1}{2|\N^{+}(v)|}} =\alpha <1.
\end{align*}
According to \cite{horn2012matrix}, $\left(\IM-{\alpha}\cdot \frac{1}{2}{\EM}^{\top}\DM^{-1}\EM\right)^{-1}$ exists and can be expanded as an infinite series
\begin{small}
\begin{equation}\label{eq:nms}
\left(\IM - {\alpha}\cdot \frac{1}{2}{\EM}^{\top}\DM^{-1}\EM\right)^{-1} = \sum_{\ell=0}^{\infty}{\alpha^\ell\left(\frac{1}{2}{\EM}^{\top}\DM^{-1}\EM\right)^{\ell}}.
\end{equation}
\end{small}
Combining Eq. \eqref{eq:z-nms} and Eq. \eqref{eq:nms} leads to Eq. \eqref{eq:zm}.
\end{proof}

\begin{proof}[\bf Proof of Lemma \ref{lem:ERWR}]
As per the definitions of $\EM$ and $\DM$, we have $\frac{1}{|\N^{+}(u_j)|}=(\DM^{-1}\EM)[u_j,e_j]$. Hence, the probability of jumping from $e_i=(u_i,v_i)$ to $e_j=(u_j,v_j)$ via $u_j=u_i$ or $u_j=v_i$ with one step in an ERW is $\sum_{u_j\in e_i}{\frac{\EM^{\top}[e_i,u_j]}{\sum_{v\in V}{\EM^{\top}[e_i,v]}}\cdot (\DM^{-1}\EM)[u_j,e_j]}=\left(\frac{1}{2}{\EM}^{\top}\DM^{-1}\EM\right)[e_i,e_j]$.
As such, the probability of an ERW originating from $e_s$ visiting $e_t$ at $\ell$-th step is $\alpha^{\ell}\left(\frac{1}{2}{\EM}^{\top}\DM^{-1}\EM\right)^{\ell}[e_s,e_t].$
Recall that at each step, the walk terminates at its current edge with a probability of $1-\alpha$. Then, the probability of an ERW originating from $e_s$ ending at $e_t$ at $\ell$-th step is $(1-\alpha)\alpha^{\ell}\left(\frac{1}{2}{\EM}^{\top}\DM^{-1}\EM\right)^{\ell}[e_s,e_t]$. By considering all the steps that the walk might stop at, we can obtain the ERW score $r(e_i,e_j)$ of two edges $e_i,e_j$ as $$r(e_i,e_j)=\sum_{\ell=0}^{\infty}{(1-\alpha)\alpha^{\ell}{\left(\frac{1}{2}{\EM}^{\top}\DM^{-1}\EM\right)}^{\ell}}[e_i,e_j]=\BM[e_i,e_j].$$
It can be observed that $r(e_i,e_j)=r(e_j,e_i)$ since ${\EM}^{\top}\DM^{-1}\EM$ is a symmetric matrix. By Eq. \eqref{eq:zm} and the above equations, we can derive that 
$\zm[e_i] =\sum_{e_j\in \EDG}{\frac{r(e_i,e_j)}{\sqrt{|\N^{+}(u_j)|+|\N^{+}(v_j)|}}}\\
 =\sum_{e_j\in \EDG}{\frac{r(e_j,e_i)}{\sqrt{|\N^{+}(u_j)|+|\N^{+}(v_j)|}}}$.
It is easy to verify that $\sum_{e_i\in \EDG}{r(e_i,e_j)}=1$. Hence, we have 
$\zm[e_i]\le \underset{{e_i\in \EDG}}{\min}{\frac{1}{\sqrt{|\N^{+}(u_i)|+|\N^{+}(v_i)|}}}$ and $\zm[e_i]\ge \underset{{e_i\in \EDG}}{\max}{\frac{1}{\sqrt{|\N^{+}(u_i)|+|\N^{+}(v_i)|}}}$, which seals the proof.
\end{proof}

\begin{proof}[\bf Proof of Lemma \ref{lem:2property}]
First, we prove that $\frac{1}{2}{\EM}^{\top}\DM^{-1}\EM$ is a semi-positive row-stochastic matrix. Consider any edge $e_i$ in $\EDG$. We have
\begin{small}
\begin{align*}
& \sum_{e_j\in \EDG}{\left(\frac{1}{2}{\EM}^{\top}\DM^{-1}\EM\right)[e_i,e_j]} \\
& = \sum_{e_j\in \EDG}\sum_{v\in V}\frac{1}{2}{\EM}^{\top}[e_i,v]\cdot (\DM^{-1}\EM)[v,e_j] \\
& = \sum_{v\in V}\frac{1}{2}{\EM}^{\top}[e_i,v] \sum_{e_j\in \EDG}(\DM^{-1}\EM)[v,e_j] = \sum_{v\in V}\frac{1}{2}{\EM}^{\top}[e_i,v] = 1.
\end{align*}
\end{small}
Since all the $m$ non-zero entries in $\frac{1}{2}{\EM}^{\top}\DM^{-1}\EM$ are positive, $\frac{1}{2}{\EM}^{\top}\DM^{-1}\EM$ is a semi-positive row-stochastic matrix. Due to its row-stochasticity, we can generalize that ${\left(\frac{1}{2}{\EM}^{\top}\DM^{-1}\EM\right)}^{\ell}\ \forall{\ell\ge 1}$ is also a semi-positive row-stochastic matrix.
Moreover, notice that $\frac{1}{2}{\EM}^{\top}\DM^{-1}\EM$ can be represented by $\frac{1}{\sqrt{2}}{\EM}^{\top}\DM^{-\frac{1}{2}}\cdot (\frac{1}{\sqrt{2}}{\EM}^{\top}\DM^{-\frac{1}{2}})^{\top}$, indicating that $\frac{1}{2}{\EM}^{\top}\DM^{-1}\EM$ and ${\left(\frac{1}{2}{\EM}^{\top}\DM^{-1}\EM\right)}^{\ell}$ are symmetric, and thus, are semi-positive bistochastic stochastic.

Let $\BM=\sum_{\ell=0}^{\infty}{(1-\alpha)\alpha^{\ell}{\left(\frac{1}{2}{\EM}^{\top}\DM^{-1}\EM\right)}^{\ell}}$. Akin to the above analysis, $\BM$ is also a symmetric matrix. Besides, for every $e_i$-th row,
\begin{align*}
\sum_{e_j\in \EDG}\BM[e_i,e_j] &= {\sum_{\ell=0}^{\infty}{(1-\alpha)\alpha^{\ell}\sum_{e_j\in \EDG}{\left(\frac{1}{2}{\EM}^{\top}\DM^{-1}\EM\right)}^{\ell}}[e_i,e_j]} \\
& = \sum_{\ell=0}^{\infty}{(1-\alpha)\alpha^{\ell}} = 1.
\end{align*}
We can conclude that $\BM$ is symmetric and row-stochastic, i.e., a semi-positive bistochastic matrix.
\eat{
Next, we prove Property 2. Notice that $\frac{1}{2}{\EM}^{\top}$ can be represented by $\DM_E^{-1}\EM^{\top}$ where $\DM_E$ is a diagonal matrix and each entry $\DM_E[e_i,e_i]=\sum_{v\in V}{\EM[v,e_i]}$, which equals to $1$ and $21$ when $\G$ is directed and undirected, respectively. For any integer $\ell\ge 1$, ${\left(\frac{1}{2}{\EM}^{\top}\DM^{-1}\EM\right)}^{\ell}\DM_E^{-1}={\left(\DM_E^{-1}\EM^{\top}\DM^{-1}\EM\right)}^{\ell}\DM_E^{-1}$ is a symmetric matrix, meaning that $${\left(\frac{1}{2}{\EM}^{\top}\DM^{-1}\EM\right)}^{\ell}[e_i,e_j]\cdot \DM^{-1}_E[e_j,e_j] = {\left(\frac{1}{2}{\EM}^{\top}\DM^{-1}\EM\right)}^{\ell}[e_j,e_i]\cdot \DM^{-1}_E[e_i,e_i].$$ Therefore, the second property is proven.
}
\end{proof}

\eat{
\begin{proof}[\bf Proof of Lemma \ref{lem:range}]
By Property 2 in Lemma \ref{lem:2property}, for any two edges $e_i,e_j\in \EDG$, $ r(e_i,e_j)= r(e_j,e_i)$. Hence, 
\begin{align*}
C_{\text{\algoabbr}}(e) &=\sum_{e_i=(u_i,v_i)\in \EDG}{\frac{ r(e_i,e)}{\sqrt{|\N^{+}(u_i)|+|\N^{+}(v_i)|}}}\\
& =\sum_{e_i=(u_i,v_i)\in \EDG}{\frac{ r(e,e_i)}{\sqrt{|\N^{+}(u_i)|+|\N^{+}(v_i)|}}},
\end{align*}
leading to $$C_{\text{\algoabbr}}(e)\le \underset{{e\in \EDG}}{\min}{\sqrt{|\N^{+}(u_i)|+|\N^{+}(v_i)|}}\cdot \sum_{e_i\in \EDG}{r(e,e_i)}$$ and $C_{\text{\algoabbr}}(e)\ge \underset{{e\in \EDG}}{\max}{\sqrt{|\N^{+}(u_i)|+|\N^{+}(v_i)|}}\cdot \sum_{e_i\in \EDG}{r(e,e_i)}$. Next, by Eq. \eqref{eq:ERWR} and Property 1 in  Lemma \ref{lem:2property}, we can get $\sum_{e_i\in \EDG}{r(e,e_i)}=1$, which seals the proof.
\end{proof}
}

\begin{proof}[\bf Proof of Theorem \ref{lem:approx-err}]
First, we denote a truncated version of $r(e_i,e)$ by $r^{\prime}(e_i,e)=\sum_{\ell=0}^{t}{(1-\alpha)\alpha^{\ell}\cdot {\left(\frac{1}{2}{\EM}^{\top}\DM^{-1}\EM\right)}^{\ell}}[e_i,e]$.
By induction, it can be shown that $\zm^{\prime}$ obtained after $t$ iterations can be represented by
$\zm^{\prime} = \xm\sum_{\ell=0}^{t}{\alpha^{\ell}\cdot {\left(\frac{1}{2}{\EM}^{\top}\DM^{-1}\EM\right)}^{\ell}}.$
Hence, we have
\begin{align*}
\zm^{\prime}[e]=\sum_{e_i\in \EDG}{r^{\prime}(e,e_i)\cdot \xm[e_i]} = \sum_{e_i\in \EDG}{\frac{ r^{\prime}(e,e_i)}{\sqrt{|\N^{+}(u_i)|+|\N^{+}(v_i)|}}}.
\end{align*}
By the definition of exact \algo in Eq. \eqref{eq:zm}, we obtain
\begin{align*}
& \zm[e]-\zm^{\prime}[e] = \sum_{e_i\in \EDG}{\frac{ (r(e,e_i)-r^{\prime}(e,e_i)) }{\sqrt{|\N^{+}(u_i)|+|\N^{+}(v_i)|}}} \\
& = \sum_{e_i\in \EDG}{\frac{  \sum_{\ell=t+1}^{\infty}{(1-\alpha)\alpha^{\ell}\cdot {\left(\frac{1}{2}{\EM}^{\top}\DM^{-1}\EM\right)}^{\ell}}[e,e_i] }{\sqrt{|\N^{+}(u_i)|+|\N^{+}(v_i)|}}} \\
& \le \frac{ \sum_{\ell=t+1}^{\infty}{(1-\alpha)\alpha^{\ell} \sum_{e_i\in \EDG}{\left(\frac{1}{2}{\EM}^{\top}\DM^{-1}\EM\right)}^{\ell}}[e,e_i]}{\underset{{(u_i,v_i)\in \EDG}}{\min}{\sqrt{|\N^{+}(u_i)|+|\N^{+}(v_i)|}}}\\
& = \frac{ \sum_{\ell=t+1}^{\infty}{(1-\alpha)\alpha^{\ell}}}{\underset{{(u_i,v_i)\in \EDG}}{\min}{\sqrt{|\N^{+}(u_i)|+|\N^{+}(v_i)|}}}.
\end{align*}
Further, using the fact $\sum_{\ell=t+1}^{\infty}{(1-\alpha)\alpha^{\ell}}=1-\sum_{\ell=0}^{t}{(1-\alpha)\alpha^{\ell}}=\alpha^{t+1}=\epsilon$, we have $\zm[e]-\zm^{\prime}[e] \le \frac{ \epsilon}{\underset{{(u_i,v_i)\in \EDG}}{\min}{\sqrt{|\N^{+}(u_i)|+|\N^{+}(v_i)|}}}$. Given that $u_i$ and $v_i$ are two endpoints of $e_i=(u_i,v_i)$, $|\N^{+}(u_i)|\ge 1$ and $|\N^{+}(v_i)|\ge 0$, Eq. \eqref{cc-abs-err} naturally follows.
\end{proof}

\begin{proof}[\bf Proof of Theorem \ref{lem:algo2}]
Suppose that Algo. \ref{alg:algo2} terminates at the beginning of $T$-th iteration. We denote by $\thetam^{(T)}$ the vector $\thetam$ at the beginning of $T$-th iteration. Notice that the vector $\thetam^{(T)}$ satisfies $\forall{e_i\in \EDG}$ $\thetam^{(T)}[e_i]\le \epsilon$. According to Lines 5-6 in Algo. \ref{alg:algo2}, we can derive that
\begin{small}
\begin{equation*}
\begin{aligned}
\zm^{\prime} = \sum_{\ell=0}^{T-1}{(1-\alpha)\alpha^{\ell}{\left(\frac{1}{2}{\EM}^{\top}\DM^{-1}\EM\right)}^{\ell}}\xm\ \\ 
\thetam^{(T)}=\alpha^{T}{\left(\frac{1}{2}{\EM}^{\top}\DM^{-1}\EM\right)}^{T}\xm.
\end{aligned}
\end{equation*}
\end{small}
On this basis,
\begin{small}
\begin{align*}
\zm - \zm^{\prime} & = \sum_{\ell=T}^{\infty}{(1-\alpha)\alpha^{\ell}{\left(\frac{1}{2}{\EM}^{\top}\DM^{-1}\EM\right)}^{\ell}}\xm\\
&= \sum_{\ell=0}^{\infty}{(1-\alpha)\alpha^{\ell}{\left(\frac{1}{2}{\EM}^{\top}\DM^{-1}\EM\right)}^{\ell}} \alpha^T{\left(\frac{1}{2}{\EM}^{\top}\DM^{-1}\EM\right)}^{T}\xm\\
&=\sum_{\ell=0}^{\infty}{(1-\alpha)\alpha^{\ell}{\left(\frac{1}{2}{\EM}^{\top}\DM^{-1}\EM\right)}^{\ell}}\cdot \thetam^{(T)}.
\end{align*}
\end{small}
Let $\BM=\sum_{\ell=0}^{\infty}{(1-\alpha)\alpha^{\ell}{\left(\frac{1}{2}{\EM}^{\top}\DM^{-1}\EM\right)}^{\ell}}$ and recall that as stated in Lemma \ref{lem:2property} $\BM$ is bistochastic, i.e., $\BM\mathbf{1}=\mathbf{1}$. Then, for every edge $e_i\in \EDG$,
\begin{align*}
\zm[e_i] - \zm^{\prime}[e_i] = \sum_{e_j\in \EDG}{\BM[e_i,e_j]\cdot \thetam^{(T)}[e_j]} \le \max_{e_j\in \EDG}{\thetam^{(T)}[e_j]}\le \epsilon,
\end{align*}
which completes the proof.
\end{proof}
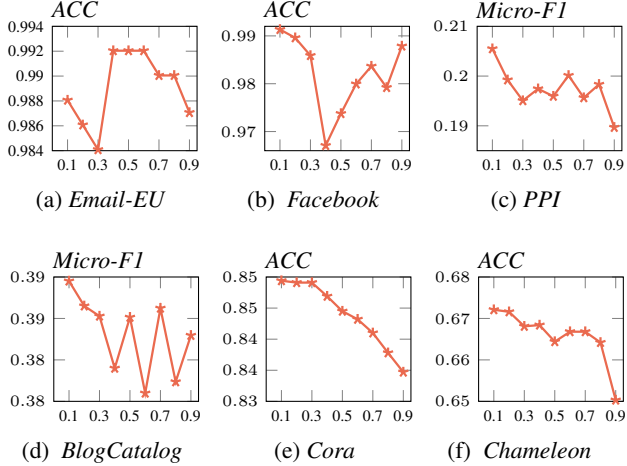
\begin{figure}[!t]
\centering
\begin{small}
\subfloat[{\em Email-EU}]{
\begin{tikzpicture}[scale=1,every mark/.append style={mark size=2pt}]
    \begin{axis}[
        height=\columnwidth/2.6,
        width=\columnwidth/2.4,
        ylabel={\it ACC},
        xmin=0, xmax=9.5,
        ymin=0.984, ymax=0.994,
        xtick={1,3,5,7,9},
        xticklabel style = {font=\tiny},
        yticklabel style = {font=\tiny},
        xticklabels={0.1,0.3,0.5,0.7,0.9},
        ytick={0.984,0.986,0.988,0.99,0.992,0.994},
        yticklabels={0.984,0.986,0.988,0.99,0.992,0.994},
        every axis y label/.style={font=\small,at={(current axis.north west)},right=3mm,above=0mm},
        legend style={fill=none,font=\small,at={(0.02,0.99)},anchor=north west,draw=none},
    ]

        \addplot[line width=0.3mm, color=EOCH-color,mark=star]  
        plot coordinates {
(1,	0.98806	)
(2,	0.98607	)
(3,	0.98408	)
(4,	0.99204	)
(5,	0.99204	)
(6,	0.99204	)
(7,	0.99005	)
(8,	0.99005	)
(9,	0.987065	)

    };
    \end{axis}
\end{tikzpicture}\hspace{0mm}\label{fig:alpha-email}%
}%
\subfloat[ {\em Facebook} ]{
\begin{tikzpicture}[scale=1,every mark/.append style={mark size=2pt}]
    \begin{axis}[
        height=\columnwidth/2.6,
        width=\columnwidth/2.4,
        ylabel={\it ACC},
        xmin=0, xmax=9.5,
        ymin=0.966, ymax=0.992,
        xtick={1,3,5,7,9},
        xticklabel style = {font=\tiny},
        yticklabel style = {font=\tiny},
        xticklabels={0.1,0.3,0.5,0.7,0.9},
        every axis y label/.style={font=\small,at={(current axis.north west)},right=3mm,above=0mm},
        legend style={fill=none,font=\small,at={(0.02,0.99)},anchor=north west,draw=none},
    ]

    \addplot[line width=0.3mm, color=EOCH-color,mark=star]  
        plot coordinates {
(1,	0.991334	)
(2,	0.989601	)
(3,	0.985888	)
(4,	0.967071	)
(5,	0.973756	)
(6,	0.979946	)
(7,	0.983659	)
(8,	0.979203	)
(9,	0.987868	)

    };
    \end{axis}
\end{tikzpicture}\hspace{0mm}\label{fig:alpha-fb}%
}%
\subfloat[{\em PPI}]{
\begin{tikzpicture}[scale=1,every mark/.append style={mark size=2pt}]
    \begin{axis}[
        height=\columnwidth/2.6,
        width=\columnwidth/2.4,
        ylabel={\it Micro-F1},
        xmin=0, xmax=9.5,
        ymin=0.185, ymax=0.21,
        xtick={1,3,5,7,9},
        xticklabel style = {font=\tiny},
        yticklabel style = {font=\tiny},
        xticklabels={0.1,0.3,0.5,0.7,0.9},
        every axis y label/.style={font=\small,at={(current axis.north west)},right=6mm,above=0mm},
        legend style={fill=none,font=\small,at={(0.02,0.99)},anchor=north west,draw=none},
    ]

        \addplot[line width=0.3mm, color=EOCH-color,mark=star]  
        plot coordinates {
(1,	0.205487623	)
(2,	0.199224575	)
(3,	0.1950492097	)
(4,	0.1974351327	)
(5,	0.1959439308	)
(6,	0.2001192962	)
(7,	0.1956456904	)
(8,	0.1983298539	)
(9,	0.1896808828	)

    };
    \end{axis}
\end{tikzpicture}\hspace{0mm}\label{fig:alpha-ppi}%
}%

\subfloat[ {\em BlogCatalog} ]{
\begin{tikzpicture}[scale=1,every mark/.append style={mark size=2pt}]
    \begin{axis}[
        height=\columnwidth/2.6,
        width=\columnwidth/2.4,
        ylabel={\it Micro-F1},
        xmin=0, xmax=9.5,
        ymin=0.375, ymax=0.39,
        xtick={1,3,5,7,9},
        xticklabel style = {font=\tiny},
        yticklabel style = {font=\tiny},
        xticklabels={0.1,0.3,0.5,0.7,0.9},
        every axis y label/.style={font=\small,at={(current axis.north west)},right=6mm,above=0mm},
        legend style={fill=none,font=\small,at={(0.02,0.99)},anchor=north west,draw=none},
    ]

    \addplot[line width=0.3mm, color=EOCH-color,mark=star]  
        plot coordinates {
(1,	0.3895110626	)
(2,	0.386506419	)
(3,	0.3852772467	)
(4,	0.3789948102	)
(5,	0.3851406719	)
(6,	0.3759901666	)
(7,	0.3862332696	)
(8,	0.3773559137	)
(9,	0.3829554766	)

    };
    \end{axis}
\end{tikzpicture}\hspace{0mm}\label{fig:alpha-blog}%
}%
\subfloat[{\em Cora}]{
\begin{tikzpicture}[scale=1,every mark/.append style={mark size=2pt}]
    \begin{axis}[
        height=\columnwidth/2.6,
        width=\columnwidth/2.4,
        ylabel={\it ACC},
        xmin=0, xmax=9.5,
        ymin=0.83, ymax=0.85,
        xtick={1,3,5,7,9},
        xticklabel style = {font=\tiny},
        yticklabel style = {font=\tiny},
        xticklabels={0.1,0.3,0.5,0.7,0.9},
        every axis y label/.style={font=\small,at={(current axis.north west)},right=3mm,above=0mm},
        legend style={fill=none,font=\small,at={(0.02,0.99)},anchor=north west,draw=none},
    ]

        \addplot[line width=0.3mm, color=EOCH-color,mark=star]  
        plot coordinates {
(1,	0.8494	)
(2,	0.8491	)
(3,	0.8491	)
(4,	0.8469	)
(5,	0.8445	)
(6,	0.8432	)
(7,	0.841	)
(8,	0.8378	)
(9,	0.8347	)

    };
    \end{axis}
\end{tikzpicture}\hspace{0mm}\label{fig:alpha-cora}%
}%
\subfloat[ {\em Chameleon} ]{
\begin{tikzpicture}[scale=1,every mark/.append style={mark size=2pt}]
    \begin{axis}[
        height=\columnwidth/2.6,
        width=\columnwidth/2.4,
        ylabel={\it ACC},
        xmin=0, xmax=9.5,
        ymin=0.65, ymax=0.68,
        xtick={1,3,5,7,9},
        xticklabel style = {font=\tiny},
        yticklabel style = {font=\tiny},
        xticklabels={0.1,0.3,0.5,0.7,0.9},
        every axis y label/.style={font=\small,at={(current axis.north west)},right=3mm,above=0mm},
        legend style={fill=none,font=\small,at={(0.02,0.99)},anchor=north west,draw=none},
    ]

    \addplot[line width=0.3mm, color=EOCH-color,mark=star]  
        plot coordinates {
(1,	0.6721	)
(2,	0.6716	)
(3,	0.6681	)
(4,	0.6684	)
(5,	0.6644	)
(6,	0.6668	)
(7,	0.6668	)
(8,	0.6642	)
(9,	0.6503	)

    };
    \end{axis}
\end{tikzpicture}\hspace{-2mm}\label{fig:alpha-chameleon}%
}%
\end{small}
 \vspace{-3mm}
\caption{The performance of \algo in downstream tasks when varying $\alpha$.} \label{fig:alpha}
\vspace{-2mm}
\end{figure}

\section{Additional Experiments}

\stitle{Environment}
All the experiments were conducted on a Linux machine powered by 2 Xeon Gold 6330
@2.0GHz CPUs and 1TB RAM. 
All the proposed methods were implemented in Python. For Edge Betweenness, we employ the NetworkX package \cite{hagberg2008exploring} for computation.

\stitle{Datasets} We experiment with six real networks whose statistics are listed in Table \ref{tab:datasets}. {\em Email-EU} \cite{leskovec2007graph} includes emails between members of a large research institution in Europe, whereas the {\em Facebook} dataset \cite{leskovec2012learning} contains participated Facebook users and their friendships, both of which are collected from SNAP \cite{snapnets}. As for the {\em PPI} and {\em BlogCatalog}, they are two popular datasets used in {\em network embedding} (NE) \cite{grover2016node2vec}. The {\em PPI} dataset is a subgraph of the protein-to-protein interaction network for Homo Sapiens, wherein node labels are obtained from the hallmark gene sets and represent biological states. {\em BlogCatalog} is a network of social relationships among the bloggers on the BlogCatalog website, where node labels correspond to the interests of bloggers.
Moreover, we include two well-known benchmark datasets for {\em graph neural networks} (GNNs), i.e., {\em Cora} \cite{yang2016revisiting} and {\em Chameleon} \cite{rozemberczki2021multi}.
We download and preprocess them using the PyTorch Geometric Library \cite{Fey/Lenssen/2019}.

\subsection{Effects of $\alpha$}
In this set of experiments, we study the effects of jumping probability $\alpha$ in \algo on the three downstream tasks when $\rho$ is fixed as $50\%$. Fig. \ref{fig:alpha} plots the spectral clustering accuracies, micro-F1 scores for node classification using node2vec, and the classification accuracy performance using GCN on six datasets when $\alpha$ is varied from $0.1$ to $0.9$. As reported, on most datasets, the best clustering and classification performance is attained when $\alpha=0.1$. The only exception is {\em Email-EU}, where the highest clustering quality is achieved when $\alpha=0.5$. Recall that $\alpha$ stands for the weight for our second optimization objective in Eq. \eqref{eq:obj}, which is also interpreted as the probability of jumping to adjacent edges in Section \ref{sec:inter-rw}.
Accordingly, a low $\alpha$ renders 
an ERW tends to stop at the edges in the vicinity of the source edge. 
The empirical results from Fig. \ref{fig:alpha} indicate that the structural information from proximal edges (direct neighbors) is more critical for node clustering and classification tasks in networks.


\end{document}